\documentclass{emulateapj}
\usepackage{apjfonts}

\def\hi{\ifmmode {\rm H}\,{\sc i}~ \else H\,{\sc i}~\fi}

\def\brizjk {BRi^\prime z^\prime JK}
\def\zphot {z_{\rm phot}}

\slugcomment{Accepted to The Astrophysical Journal}

\shorttitle{Quiescent galaxies in a bicolor sequence}
\shortauthors{Williams et al.}

\begin{document}

\title{Detection of quiescent galaxies in a bicolor sequence from $z=0-2$
\altaffilmark{1}}


\author{Rik J. Williams\altaffilmark{2}, 
        Ryan F. Quadri\altaffilmark{2},
	Marijn Franx\altaffilmark{2},
        Pieter van Dokkum\altaffilmark{3},
	Ivo Labb\'e\altaffilmark{4}}
\altaffiltext{1}{
Based in part on data collected at Subaru Telescope through the ``Subaru
Observatory Project,'' which is operated by the National Astronomical 
Observatory of Japan.
}
\altaffiltext{2}{Leiden Observatory, Leiden University, Niels Bohrweg 2,
                 NL--2333 CA  Leiden, The Netherlands}
\altaffiltext{3}{Department of Astronomy, Yale University,
                 New Haven, CT 06520--8101}
\altaffiltext{4}{Carnegie Observatories, 813 Santa Barbara Street,
Pasadena, CA 91101; Hubble Fellow}

\email{williams@strw.leidenuniv.nl}

\begin{abstract}
We investigate the properties of quiescent and star--forming galaxy
populations to $z\sim 2$ with purely photometric data, employing a novel
rest--frame color selection technique.  From the UKIDSS 
Ultra--Deep Survey Data Release 1, with matched optical and mid--IR photometry
taken from the Subaru XMM Deep Survey and Spitzer Wide--Area Infrared
Extragalactic Survey respectively, we construct a $K$--selected galaxy
catalog and calculate photometric redshifts.  Excluding stars,
objects with uncertain $\zphot$ solutions, those that fall in bad or 
incomplete survey regions, and those for which reliable rest--frame colors 
could not be derived, 30108 galaxies with $K<22.4$ (AB) and $\zphot\le 2.5$ 
remain.  The galaxies in this sample are found to occupy two distinct 
populations in the rest--frame $U-V$ vs.~$V-J$ color space: a clump of 
red, quiescent galaxies (analogous to the red sequence) and a
track of star--forming galaxies extending from blue to red $U-V$ colors.
This bimodal behavior is seen up to $z\sim 2$.  Due to a combination
of measurement errors and passive evolution, the color--color diagram
is not suitable to distinguish the galaxy bimodality at $z>2$ for this sample, but
we show that MIPS 24$\mu$m data suggest that a significant population of
quiescent galaxies exists even at these higher redshifts.  At $z=1-2$,
the most luminous objects in the sample are divided roughly equally between
star--forming and quiescent galaxies, while at lower redshifts most of
the brightest galaxies are quiescent.  Moreover, 
quiescent galaxies at these redshifts are clustered more strongly than
those actively forming stars, indicating
that galaxies with early--quenched star formation may occupy more massive 
host dark matter halos.  This suggests that the end of star formation is 
associated with, and perhaps brought about by, a mechanism related to 
halo mass.
\end{abstract}



\section{Introduction}
The stellar mass in the local universe lies largely within two distinct
classes of galaxies: actively star--forming spirals, and massive
ellipticals with evolved stellar populations and little current
star formation.  These populations show strongly bimodal behavior
in a number of measured and derived quantities, including color
\citep{baldry04}, the 4000\AA--break strength $D_n(4000)$
\citep[an indicator of stellar population age;][]{kauffmann03}, and 
clustering strength \citep{budavari03}.  
Relatively low--mass galaxies with high
star--formation rates have also been found at high redshifts
\citep[e.g., $z>3$ Lyman--break galaxies or LBGs;][]{steidel96}.
Surprisingly, the population of massive, ``dead'' galaxies appears to
persist to high redshift as well \citep[e.g.][]{labbe05,daddi05,kriek06,kriek08b}, even though
the universe was only a few Gyr old at that point.  Since the stellar
populations of evolved galaxies often have ages that are a significant
fraction of the age of the universe at these redshifts,
the implication is that they must have formed (and
their star formation ceased) far earlier than anticipated by standard
hierarchical structure formation models.

Investigating the evolution of star formation activity and the assembly 
of stellar mass
over the past $\sim 12$\,Gyr are therefore two key avenues to understanding
how the present galaxy population came to be, and large samples of galaxies
during this peak formation epoch ($z\sim 2$) are needed for this.
As spectroscopy of faint
objects is extremely expensive with regard to telescope time, several
highly efficient broadband selection techniques have been designed to weed out 
low--redshift interlopers.  In addition to the 
aforementioned LBG selection, which primarily finds unobscured
star--forming galaxies at $z>3$, the distant red galaxy (DRG) criterion 
\citep[$J-K>2.3$ (Vega) or $>1.34$ (AB);][]{franx03} instead tends to find 
the most 
{\it massive} galaxies at $2<z<3$ \citep{vdokkum06}, about half of
which show signs of heavily--obscured star formation \citep{papovich06}.  
Other methods, such as the $BzK$ selection of \citet{daddi04}, are adept at 
selecting $z\sim 2$ galaxies over a wide range of masses and star--formation 
rates.

Each of these techniques has its advantages and disadvantages, but one point
has become clear: to effectively study the mass evolution of high--redshift
galaxies, deep near--infrared data are crucial.  The observed $JHK$ bands 
trace rest--frame optical light at $z\sim 2-3$ and provide a more reliable
indicator of stellar mass than the rest--frame UV (observed optical); indeed,
DRGs themselves appear to represent an important fraction, if not the
majority, of stellar mass at high redshifts \citep{rudnick06,grazian07}.
Such galaxies are typically faint in the observed optical bands,
either from old stellar populations or dust obscuration, and are 
often missed by purely optical selection techniques \citep{quadri07b}.
Selecting galaxies from deep near--infrared fields is therefore likely
to yield a more complete picture of the stellar mass at high redshift than
similar studies at optical wavelengths.

Until recently, the small sizes (at most a few square arcminutes) and
relatively low efficiencies of the available infrared detectors 
made large, deep near--IR surveys impractical.  Such projects therefore 
typically followed the ``pencil--beam'' approach, surveying a single frame
to high sensitivity \citep[e.g., the {\it Faint Infrared
Extragalactic Survey, FIRES};][]{franx03}, or alternatively observing 
somewhat larger areas at the expense of depth \citep[e.g., 
GOODS--S;][]{giavalisco04}.  While such surveys have
revealed a great deal about the high--redshift universe, the necessarily
small survey volumes proved problematic for statistical studies 
of galaxy populations (as a result of both cosmic variance and small--number 
statistics).  

Large--format  near--IR detectors on 4m--class telescopes (such as 
ISPI, WFCAM,
and the upcoming VISTA camera) now provide the ability to conduct deep
surveys over wide areas, and a number of past, present, and planned 
projects take advantage of this capability -- examples include
the Multiwavelength Survey by Yale--Chile \citep[MUSYC;][]{gawiser06,quadri07b},
the UKIRT Infrared Deep Sky Survey \citep[UKIDSS;][]{lawrence07}, and
ULTRA--VISTA.  Until this latter survey is complete, the UKIDSS 
Ultra--Deep Survey 
(UDS\footnote{\url{http://www.nottingham.ac.uk/astronomy/UDS/}}) is 
the premier near--IR dataset in
terms of depth and area.  Overlapping deep optical imaging from the
Subaru--XMM Deep Survey \citep[SXDS;][]{sekiguchi04} and shallow mid--IR data 
from the
{\it Spitzer} Wide--Area Infrared Extragalactic 
\citep[SWIRE;][]{lonsdale03} survey provide complementary data over
a broad range of wavelengths.

Here we derive a multiband $K$--selected galaxy catalog from the
UDS, SXDS, and SWIRE data, calculating photometric redshifts and rest--frame
colors for all objects in the overlapping survey area.  Rather than rely 
on a specific color selection 
technique, we instead define samples of galaxies at various redshifts directly 
from the computed $\zphot$ values, and further subdivide the samples into 
star--forming and quiescent galaxies based on their rest--frame colors.
The color evolution with redshift, and the clustering of $1<\zphot<2$
galaxies, are then determined.  Readers who are primarily interested
in these science results can find this discussion beginning
in \S\ref{sec_analysis}.  

We present an overview of the
survey data in \S\ref{sec_surveys}, and the process of
preparing and matching the datasets to each other
in \S\ref{sec_data}.  The extraction of the catalog is then
discussed in \S\ref{sec_catalog}, and the derivation of photometric
redshifts and rest--frame colors in \S\ref{sec_photz}.  \S\ref{sec_analysis}
describes the bimodality of galaxies in rest--frame color space and
the separation into quiescent and star--forming samples, and the
clustering properties of these two populations are discussed in 
\S\ref{sec_clustering}.
Further discussion, caveats, and a summary of the results can then be 
found in \S\ref{sec_discussion}, \S\ref{sec_caveats}, and 
\S\ref{sec_conclusions} respectively.

A cosmology with
$H_0=70$\,km\,s$^{-1}$\,Mpc$^{-1}$, $\Omega_m=0.3$, and $\Omega_\Lambda=0.7$
is assumed throughout, and
magnitudes are quoted in the AB system unless otherwise noted.

\section{Data characteristics} \label{sec_surveys}
In this analysis we employ the reduced $J$-- and $K$--band UDS mosaics 
provided as a subset of the UKIDSS Data Release 1
\citep[DR1;][]{warren07a}; $H$--band data
were not available in this release.  The UKIDSS project, defined
in \citet{lawrence07}, uses the UKIRT Wide Field Camera 
\citep[WFCAM][]{casali07} and a photometric system described in
\citet{hewett06}; the pipeline processing and science archive are
described in Irwin et al. (in preparation) and \citet{hambly08} respectively.  
The UDS field consists of a single 
repeatedly--observed UKIDSS survey tile (in turn comprising four subsequent
WFCAM observations arranged in a square to produce uniform coverage), with
a total area of $0.77$\ deg$^2$ and reaching nominal depths of $K<23.45$ 
and $J<23.55$ (AB;
$5\sigma$ point--source threshold in a 2\arcsec\ aperture).  The 0\farcs4
pixel scale of WFCAM does not sufficiently sample the $K-$band PSF
of $\sim 0\farcs7$, so the individual observations are offset by 
fractions of a pixel and the combined mosaics regridded by a factor of 3
to a pixel scale of 0\farcs 134.  Note that while
Data Release 2 has since been made available \citep{warren07b}, this 
release contains
identical UDS mosaics to DR1; only data in the shallower UKIDSS surveys
were updated.  

\begin{figure}
\epsscale{1.0}
\plotone{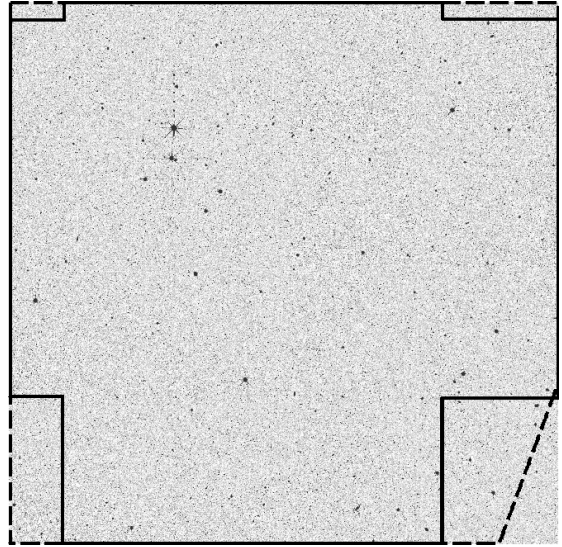}
\caption{UKIDSS $K$-band image of the UDS field, with SXDS ({\it solid line})
and SWIRE ({\it dashed line}) coverage overplotted.  The image is about
52\arcmin\ on a side; north is up and east is to the left.
\label{fig_fields}}
\end{figure}

Optical imaging of most of the UDS field is provided by the Subaru--XMM 
Deep Survey (SXDS) beta release\footnote{\url{http://www.naoj.org/Science/SubaruProject/SXDS/}},
a deep wide--field survey employing the Subaru telescope and Subaru Prime 
Focus Camera \citep[Suprime--Cam or SUP;][]{miyazaki02}.  This imager 
contains ten $2048\times 4096$
pixel CCDs arranged in a $5\times 2$ mosaic, providing a rectangular
field--of--view of roughly $34\arcmin \times 27\arcmin$.  The SXDS then
consists of five overlapping SUP fields arranged in a cross--shaped
pattern (three adjacent fields in the north--south direction, and two
on the eastern and western flanks with the camera rotated $90^\circ$;
see Figure~\ref{fig_fields}).  
Reported $5\sigma$ depths (in a 2\arcsec\ aperture) are 
$B<27.5$, $R<26.8$, $i^\prime<26.5$, and $z^\prime<25.4$, though the
sensitivity varies by $\sim 0.2$ between the five fields
(with a $\sim 0.6$ mag variation in $z$); our independent sensitivity
estimates are comparable to the reported values.

The SWIRE survey covers a total of
$\sim 49$\,deg$^2$ with IRAC and MIPS in six subfields, one
of which (the XMM--LSS field) includes nearly the entire UDS/SXDS area. 
The orientations of the IRAC/MIPS fields are at an angle to the UDS; 
therefore, only a small
triangular region in the southwest corner of the UDS has no SWIRE
coverage (shown in Figure~\ref{fig_fields}).  
SWIRE reaches nominal $5\sigma$ point--source depths of 3.7, 5.4, 48, and 
37.8 $\mu$Jy in the 
3.6, 4.5, 5.8, and 8.0\,$\mu$m IRAC bands respectively, and 230\,$\mu$Jy
in MIPS 24\,$\mu$m.
The IRAC mosaics of this field are
divided into 16 separate subimages, with four of these having some overlap
with the UDS field; the MIPS 24\,$\mu$m image of the field is
available as a single file.  The relevant subimages of all four IRAC bands,
as well as the full 24\,$\mu$m MIPS image, were
obtained from the public SWIRE data 
website\footnote{\url{http://swire.ipac.caltech.edu/swire/astronomers/}\\
\url{data\_access.html}}.  Due to their far broader point--response functions 
and lower sensitivities, we do not analyze the MIPS 70\,$\mu$m or 160\,$\mu$m 
data here; similarly, the $5.8\mu$m and $8\mu$m data are excluded 
due to their order--of--magnitude lower sensitivity compared to the first two
IRAC bands.  Furthermore, since most objects seen in the UDS $K$--band are not 
individually detected in the SWIRE $24\mu$m image, the $24\mu$m
data are therefore not included in our final $K$--selected catalog
(but a stacking analysis of these data are discussed in \S\ref{sec_mips}).
A summary of the data employed in the production of our $K$--selected
catalog, including areas and measured limiting
depths (as discussed in \S\ref{sec_complete}) is given in 
Table~\ref{tab_limits}.

\section{Data preparation} \label{sec_data}
\subsection{Image registration}
The simplest way to achieve consistent flux measurements across multiple
bands with arbitrary PSFs is to transform the images such that they are all 
on the 
same coordinate system and pixel scale.  The pixel scale of the UDS
images is 0\farcs134; however, all of these images have PSF FWHMs
larger than 0\farcs7 (the UDS $K$--band seeing).  To reduce file sizes
and processing time, we thus adopt
a standard pixel scale of 0\farcs202, the native scale of the SXDS
images.  This is somewhat coarser than the original regridded UDS pixel
scale, but still sufficiently samples all PSFs.

The SWarp software package\footnote{See 
\url{http://terapix.iap.fr/rubrique.php?id\_rubrique=49} for 
further information} \citep{bertin02} was used to crop,
resample, and transform these sets of images to a common coordinate system
before combining the subimages (in the cases of the IRAC and SXDS data) 
into single mosaics.  
Since the UDS $K$ image has the best seeing and is the band from which 
objects are detected, 
this was taken as the ``standard'' field to which all other images were
registered.  The $J$ and $K$ data were resampled to the SXDS pixel scale
and cropped to remove noisy and missing data regions near the edges, resulting
in 52\farcm4 by 53\farcm0 images.  Likewise, the 
SWIRE MIPS 24\,$\mu$m image and the IRAC subimages that cover the UDS field 
were transformed, resampled, and cropped to cover the same area.
However, since the five SXDS pointings taken with each filter can exhibit 
different
seeing and geometric distortions, those images were PSF--matched 
(see \S\ref{sec_psf}) and astrometrically corrected prior to combining
with SWarp.

As SWarp transformed the images based solely on the astrometric information 
in the
image FITS headers, some distortions are likely to remain (particularly
between data from differing instruments).
The measured positions of stars were thus used to correct these distortions.
Source Extractor version 2.5.0 \citep[SExtractor;][]{bertin96} was first 
used to find the positions
of bright objects in the $K$ image and measure their approximate $J$ and $K$ 
fluxes.  $1179$ stars were selected on the basis of their blue IR colors 
($J-K<0$) and bright but unsaturated $K$--band fluxes ($17.4<K<21.4$).  
The radial profiles of these objects were
fitted with Gaussians to determine their centers and PSF widths in the
$K$ band; all but a negligible fraction ($<1$\%)had FWHMs
comparable to the nominal image resolution, indicating that this simple 
color cut indeed primarily selects point sources.  Next, a centroid 
algorithm was employed to 
determine the positions of these stars in each of the $\brizjk/3.6/4.5$
images.  Finally, the pixel positions of these stars along
with the IRAF tasks {\tt geomap} and {\tt geotran} were used to 
transform the SXDS, UDS $J$, and IRAC images to the same coordinate system as 
the UDS $K$ mosaic.  To check the agreement between the images, the pixel 
coordinates of these stars were again measured in the transformed images
and found to agree well, e.g.~in the $B$ mosaic the image coordinates agree
with those in the $K$ image to within 0.5 pixel (0\farcs 1) in 75\% 
of cases, with a median absolute deviation of 0\farcs06.  Similar offsets
are seen in the transformed IRAC $3.6$ and $4.5\mu$m images (in addition,
automatic astrometric corrections are performed in the IRAC deblending 
described in \S\ref{sec_iracphot}).  Such 
astrometric deviations typically translate to systematic flux errors of at 
most $\sim 1$\%.

\subsection{IR/Optical PSF matching} \label{sec_psf}
The observed seeing width varies significantly between the different bands:
point sources in the UDS have FWHMs of approximately 0\farcs7 
in $K$ and 0\farcs8 in $J$, while in SXDS they range between 
$0\farcs7-0\farcs9$ depending on the field and band.  Additional
higher--order variations are likely to be present in the PSFs due to 
differences 
in the two instruments' optics.  Directly performing aperture photometry
on these images would lead to flux offsets in the different bands (with
less flux falling inside the aperture for the poorer--seeing images), and
thus systematic color errors.

To mitigate this, all images were smoothed to 
the same PSF before fluxes were measured.  Although this can be 
accomplished by convolving each with a
simple Gaussian or higher--order kernel, the multitude 
of stars within these images can themselves be used to generate
empirical kernels that contain all structural 
information about the PSFs.  This is accomplished by deconvolving
a low--resolution (``reference'') PSF with one constructed from a 
higher--resolution image:
\begin{equation}
K = P_{\rm ref} \otimes^{-1} P_{\rm hires}
\end{equation}
where $P_{\rm ref}$ and $P_{\rm hires}$ are the low--resolution and
high--resolution PSFs respectively.
The resulting kernel can then be convolved with the high--resolution 
image in its entirety to bring it to the same PSF as the reference image.
\begin{equation}
I_{\rm match}(x,y)= I_{\rm hires}(x,y) \otimes K
\end{equation}
Since the deconvolution step strongly magnifies any intrinsic
noise in the PSFs, a large number of bright (but unsaturated) stars are 
necessary in each band to produce PSFs with sufficiently low noise levels.

For the $J$ and $K$ images, the PSFs were constructed from the same stars that 
were selected
for astrometric matching, keeping the 200 brightest in each band.  However,
many of these stars are saturated in the SXDS images due to the
much longer exposure times employed in this survey.  As the effects of
saturation are far more detrimental for the PSF shape than in the 
astrometric correction, a different method for selecting stars in the 
SXDS images is necessary.
For these images, the IRAF {\tt psfmeasure} task was used to estimate 
widths of unsaturated
objects in the image based on a Gaussian fit.  In a histogram of the widths
of these objects, there is an obvious cutoff on the low--FWHM end
accompanied by a sharp peak and an overlying broad distribution
extending to larger FWHMs.  Objects in the narrow peak were presumed to
be point sources to within the image resolution, and the brightest of 
these were selected as stars in each optical image.

\begin{figure}  
\plotone{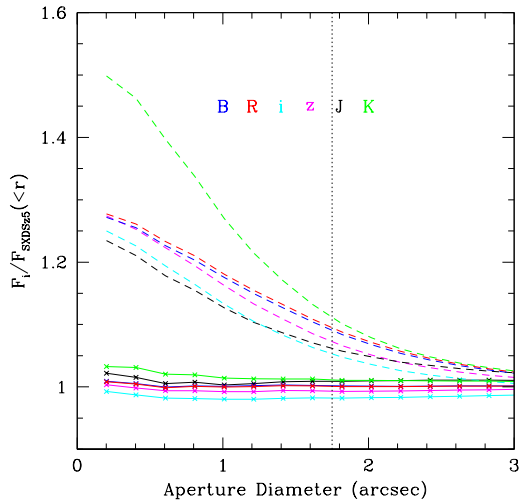}
\caption{Average growth curves of the six optical/IR bands divided by
the growth curve of
the worst image (SXDS--West $z$-band, with a PSF FWHM $\sim 0\farcs 9$).  
Relative growth curves before ({\it dashed}) and after ({\it solid with 
points}) convolution to the broadest PSF are shown.  The vertical dotted 
line shows the ``color'' aperture size of 1\farcs75.  Typical systematic
flux errors due to PSF matching at this aperture size are $<2$\%.
\label{fig_gc}}
\end{figure}

Square postage stamps (21 pixels or 4\farcs2 on a side) of the stars selected
in each image were then created, and these in turn were median--combined
to create 22 empirical PSFs (one for each of the 20 
SXDS subimages, plus one for each UDS mosaic).
Enclosed flux as a function of aperture size (growth curves) were then created
for these PSFs.  The ``slowest--growing''
growth curve, indicating the broadest PSF, is that of the western SXDS
$z-$band image with a FWHM of $\sim 0\farcs9$.  Thus, we take this as the 
reference PSF and match the other 19 SXDS fields and two UDS 
images to it.  The average growth curves for point sources in the
smoothed $\brizjk$ images, expressed as a fraction of the SXDS--West 
$z^\prime$ (reference) growth curve, before and after smoothing, are shown 
in Figure~\ref{fig_gc}.  At the aperture sizes used for flux
measurements in the following section ($>1\farcs75$ or $\sim 9$ pixels in 
diameter), the PSFs are matched to within 2\%.

The point--response functions of the IRAC 3.6\,$\mu$m and 4.5\,$\mu$m 
images are roughly $1.7\times$ as broad as
in the worst SXDS image ($\sim 1\farcs 6$ for both); 
they also contain significant non-Gaussian structure resulting from point
source diffraction.  While it would have been simpler in principle to 
smooth the $\brizjk$ images to the IRAC PSF shape, this would substantially
reduce the detection efficiency in $K$ while (for most high--redshift 
sources) providing only weak detections in IRAC to begin with.  Additionally, 
blending from nearby sources is more likely to adversely affect simple 
aperture photometry in the IRAC images.  PSF--matching and photometry are 
thus performed on these data using a different technique, described 
in \S\ref{sec_iracphot}.

\section{Catalog extraction} \label{sec_catalog}
\subsection{Source detection and photometry}\label{sec_sextractor}
To measure fluxes of sources in the $\brizjk$ images we used
SExtractor in dual--image mode, whereby one image is used to detect sources
and aperture photometry is performed on another image registered to the
identical coordinate system.  The
unsmoothed (pre--PSF matching) $K$ mosaic was used as the detection image, 
and fluxes were measured in each of the six PSF--matched IR/optical mosaics.
The sensitivity across the $K$ image is not perfectly uniform due to
overlapping exposures and array efficiency variations; such nonuniformity
can lead to corresponding variations in the number of detected galaxies,
and thus to spurious clustering signals.  We used the UKIDSS--supplied 
confidence map to construct an RMS map of the $K$ image, and this
map was applied as the weight image within SExtractor 
in order to detect sources in an effectively
noise--equalized and uniform manner.
Several initial SExtractor runs were performed with varying detection
parameters and the results checked by eye; the final choice of parameters 
appeared to find all faint objects with few spurious detections
(see \S\ref{sec_complete}).

\begin{figure}
\plotone{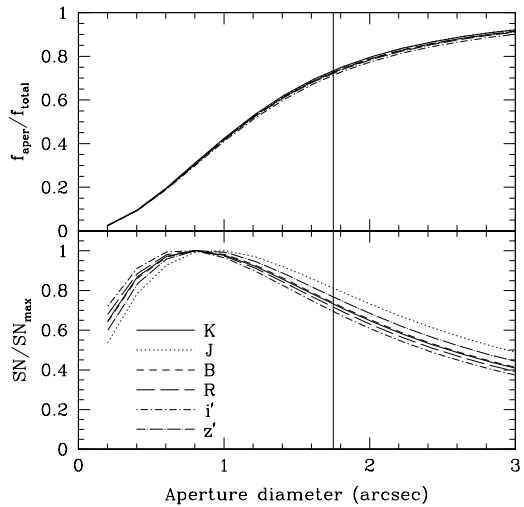}
\caption{{\it Top panel:} Median growth curves of point sources in
the PSF--matched $\brizjk$ images; {\it Bottom panel:} Signal--to--noise in the
six bands, defined as the growth curve divided by the empty aperture
noise function in each band, normalized to the peak SN in each band.
The solid vertical line denotes the color aperture size of 1\farcs75.
\label{fig_gcsn}}
\end{figure}

Total fluxes were determined using 
a flexible elliptical aperture \citep[SExtractor's 
{\tt AUTO} aperture;][]{kron80}.  To ensure that color measurements are
consistent, fluxes in a fixed circular aperture were also measured.  
The optimal size of this ``color'' aperture is subject
to two primary considerations: while a larger aperture encloses a greater 
fraction of a given object's flux, it also suffers from higher uncertainty 
due to background
fluctuations.  To determine the best aperture size, we divided the
growth curve of each image by its noise function (i.e., photometric
uncertainty versus aperture size, described in detail in
\S\ref{sec_photerr}).  This is effectively equivalent to the signal--to--noise
(SN) of each image as a function of aperture size, and is shown in the 
lower panel 
of Figure~\ref{fig_gcsn}.  

The $K$--band SN peaks at a diameter of $\sim 4$ 
pixels or 0\farcs8, which is lower than the ideal aperture size theoretically
expected with uncorrelated noise ($\sim 1\farcs 0$), but
consistent with the $1.1-1.4\times$FWHM optimal aperture found for the MUSYC 
data \citep{quadri07b}.  The growth curves shown in the top panel of
Figure~\ref{fig_gcsn} indicate 
that an aperture of this size misses about 70\% of a point source's flux. 
Such a small aperture is highly susceptible to systematic errors 
from imperfect PSF matching and astrometric offsets.  Additionally,
larger apertures appear to produce slightly bluer $J-K$ colors (on the
order of 0.05 magnitudes between 1\arcsec\ and 2\arcsec\ apertures),
perhaps due to intrinsic galaxy color gradients and/or imperfect image
matching.  To
ensure more accurate colors of non--point sources, as a compromise
we adopt a somewhat larger
``color'' aperture of 1\farcs75 diameter.  This more than doubles the enclosed
flux while only decreasing the SN by $\sim 30$\% from the optimal value.

\begin{figure}
\plotone{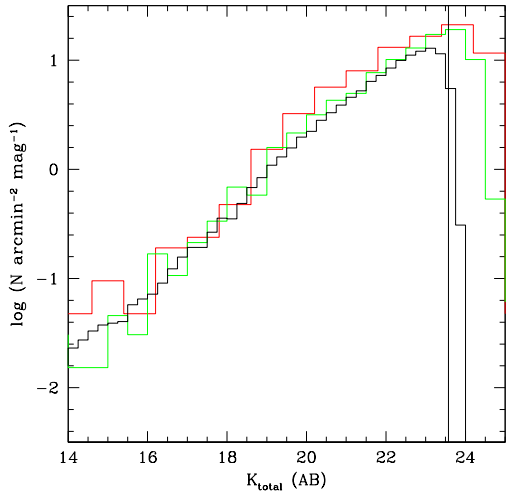}
\caption{Distribution of total $K$ magnitudes for objects in the UDS 
$K$-selected catalog detected
at $>5\sigma$ significance in the ``color'' (1\farcs 75 diameter) aperture
{\it black histogram}.
The vertical black line indicates the mean $5\sigma$ detection limit of
$K=23.57$.  Overplotted are raw number counts from the \emph{FIRES} MS1054-03
cluster field \citep[][{\it red histogram}]{forster06} and the CDFS--GOODS
$K$-selected catalog \citep[][{\it green histogram}]{wuyts08}.
\label{fig_kdist}}
\end{figure}

In cases where the {\tt AUTO} aperture is smaller than 1\farcs75
(primarily occurring for faint, point--like sources), the color aperture
flux is substituted for the total flux.  All total fluxes are corrected
for flux falling outside the aperture assuming a minimal (point--like) 
source flux distribution.
The SExtractor photometry was verified by comparing the $z^\prime$ total 
flux values with those of objects
in spatially coincident HST/ACS images.\footnote{Found using the MAST archive,
\url{http://archive.stsci.edu}.}  The only useful ACS data in the archive
were taken with the F850LP filter, which is approximately equivalent
to SXDS $z^\prime$; fluxes of objects in these ACS data were consistent
with those derived from SXDS.  The
photometric zeropoints of the SXDS $z^\prime$ data thus appear to be 
properly calibrated. Since the
same calibration techniques and standard star fields were used 
by the SXDS team to determine the
$BRi^\prime$ image zeropoints, we assume the calibrations in these bands 
are accurate as well (though the lack of available verification data should be 
kept in mind).  Figure~\ref{fig_kdist} shows the distribution of $K$ magnitudes
for all objects detected by SExtractor, with number counts overplotted from two
other $K$--selected samples: FIRES MS1054-03 \citep{forster06}
and CDFS-GOODS \citep{wuyts08}.
The number counts of the three surveys are similar to the UDS limiting
magnitude, though the enhanced number of faint objects in the FIRES
field (due to the presence of the $z=0.83$ cluster MS $1054-03$) is
evident.

\subsection{IRAC PSF--matching and photometry}\label{sec_iracphot}
As previously mentioned, the optical and NIR images were smoothed to 
the seeing of the 
SXDS--West $z^\prime$--band data rather than the much broader point--response
function of the IRAC data.  The direct aperture photometry
from SExtractor would therefore not measure IRAC fluxes that are
properly matched to the other bands.  Moreover, the lower--resolution
IRAC images are more prone to blending than the other bands, which can
introduce significant systematic errors in the  measured colors.
We thus employed the method of \citet{labbe06} to correct this, 
summarized below.

Sources that are bright in $K$ are also typically bright in the 
3.6/4.5\,$\mu$m bands; thus, 
the $K$ image is used as a high--resolution template to 
deblend the IRAC images.  
Convolution kernels used to transform from
the $K$ to the IRAC PSFs are constructed from bright and unsaturated sources
(computed by fitting a series of 
Gaussian--weighted Hermite functions to the Fourier transforms of the 
objects in IRAC and $K$), and a smoothed map of the kernel
coefficients as a function of image position is then produced.
For a given object detected in $K$, neighboring sources are
fitted and subtracted out using the local convolution kernel (derived
from the aforementioned smoothed coefficient map) and the
SExtractor--derived segmentation map.  This results in a ``cleaned'' IRAC 
subimage containing only the source in question; standard
photometry with a 3\arcsec\ diameter aperture\footnote{
As with the UDS and SXDS images, this aperture size was chosen as a 
compromise between maximal signal--to--noise and enclosing a sufficient
fraction of a point source's flux.} is then performed on the source, and
this process is repeated for each object in the $K$--selected catalog.
Through a visual inspection of the IRAC residual image with all sources
subtracted, we conclude that this method effectively removes
contaminating sources \citep[for an illustration of this technique, see 
also Figure 1 in][]{wuyts07}.

Since the IRAC photometry from this procedure is
performed using a larger aperture on images with significantly
broader PSFs than the matched $\brizjk$ images, these fluxes must be
corrected in order to obtain accurate IRAC--IR/optical colors.  This
is accomplished by multiplying the IRAC flux measured in a 3\arcsec\ 
aperture by a correction factor to obtain a ``matched'' IRAC flux:
\begin{equation}
f_{\rm IRAC,match}=f_{\rm IRAC}(3\arcsec) \times \frac{f_{\rm K}(1\farcs 75)}{f_{\rm K,conv}(3\arcsec)}
\end{equation}
where $f_{\rm K,conv}$(3\arcsec) is the flux of the object measured in the
$K$--band image after smoothing it to the IRAC resolution, and
$f_{\rm K}$ is the K--band flux measured in the fixed ``color'' aperture
(see \S\ref{sec_sextractor}).  This correction implicitly assumes
that the ratio of fluxes between the two aperture sizes is the same
for $K$ and IRAC, i.e. color gradients
between 1\farcs75 and 3\arcsec\ are insignificant.  While this technique 
could introduce
systematic errors in measurements of extended sources with strong 
color gradients, at the redshifts primarily considered here
($z\ga 1$) most galaxies have small angular sizes and 
are not likely to be adversely affected.

\subsection{Bad pixels \label{sec_badpix}}
All three sets of images (UDS, SXDS, SWIRE) contained regions where either no
data were available or the derived photometry was unreliable.  Thus,
even though SExtractor finds objects in these regions, their fluxes
should be considered unreliable. To account for this we created bad--pixel
maps for each image set; if an object's ``color'' aperture contains 
any such bad pixels, we set a flag indicating as such in the catalog.  
The bad pixel maps created for each survey are mostly a result of the
following effects:
\begin{enumerate}
\item{
UDS $JK$:{\it False sources and large negative residuals.}  
In rows and columns containing bright stars, there are often significant
positive and negative images (repeated at regular intervals) as a result
of crosstalk between WFCAM detector nodes.  These ``false positive'' images,
while obviously not true astrophysical sources upon visual inspection, 
are nonetheless picked up as
real objects by SExtractor.  To make matters worse, many such ``false 
positives'' in $K$ have negative apparent flux in $J$, causing these
to appear as very red objects.  They can thus mimic high--redshift near--IR
galaxy colors and may be an important contaminant in our catalog.
Fortunately, adjacent to most of the false positive images in $K$,
some strongly negative residuals are also typically seen.  We thus
searched for such strong $K$ residuals in the image, removing any object
falling on or within 10 pixels of any such artifact as bad.  As a secondary
check, we searched for objects which were detected in $K$ but undetected
in the sum of the $B+R+i^\prime+z^\prime$ mosaics.  This cut finds an 
additional $\sim 1900$ possible bad spots.  A visual inspection confirms
that about $80$\% of these are indeed artifacts (including diffraction
spikes from bright stars and meteor streaks in $K$) that were undetected
by the aforementioned ``negative residual'' technique, and most of the
remaining $20$\% are extremely faint and likely to be spurious detections.
However, since some of these optically--nondetected objects are real, 
they have been 
included in the catalog with a flag indicating that they are likely to
be bad.

}
\item{
SXDS $BRi^\prime z^\prime$: {\it Non--covered regions and bright stars.} 
As shown in Figure~\ref{fig_fields}, the corners of the $UDS$ fields
are not covered by the SXDS imaging.  Additionally, bright stars in SXDS
are surrounded by concentric halos and extended readout streaks.  The
Subaru team has provided files which define the regions affected by these 
bright 
stars; a visual inspection of the SXDS frames confirms that these regions
accurately describe the affected areas.  For each of the five pointings
we created a map such that pixels falling within these regions had a value
of 1, and zero otherwise.  Using SWarp, these bad pixel masks were then
transformed and combined into a single mosaic on the same coordinate system
as the other images (assigning a value of 1 to the corners with no
SXDS coverage as well).
}
\item{
SWIRE: {\it One non--covered region.} 
The SWIRE survey spans an area many times larger than the UDS field,
with nearly uniform coverage over the UDS/SXDS area except for a small 
triangular 
region in the southwest corner of the field.  As most of this triangle
is also outside the SXDS coverage area, reliable IRAC flux measurements 
(or upper limits) are available 
for essentially all objects with SXDS photometry.  Objects with coordinates
falling in this missing SWIRE piece are flagged in the catalog.
}
\end{enumerate}

It should be noted that, while these techniques were effective at 
automatically finding many false detections and areas of unreliable 
photometry, some undetected artifacts are likely to remain in the 
full catalog.  For the analysis described herein we therefore visually 
inspected subsets of our samples to ensure that such artifacts 
do not significantly affect our results.

\subsection{Photometric Error Determination}\label{sec_photerr}
The uncertainty of a flux determined through aperture photometry contains
contributions from the intrinsic Poisson error in the source counts
and from background noise.  While the first (automatically calculated by
SExtractor) is straightforward to determine  
and is the dominant source of uncertainty for bright objects,
the background uncertainty is important for fainter sources.  For perfectly
random (uncorrelated) background noise the additional photometric 
uncertainty is simply $\sigma_{\rm aper}\sim\sigma\sqrt{N}$, where $N$ is the
number of pixels contained within the aperture and $\sigma$ is the 
pixel--to--pixel RMS.  On the other extreme, if the noise is perfectly
correlated, then $\sigma_{\rm aper}\sim\sigma N$.  
Typical images will fall somewhere between these two cases,
exhibiting a general form of $\sigma_{\rm aper}\sim \sigma(\sqrt{N})^\beta$. 

We estimated the background uncertainty following the method of
\citet{labbe03} and \citet{gawiser06}.
First, to estimate this power--law index, several hundred 
apertures of a given size were randomly placed on empty parts of the images 
(i.e., those containing no emission detected by SExtractor).  
The RMS variation $\sigma_{\rm aper}$ was found by fitting a Gaussian 
to the resulting histogram of aperture fluxes, and this 
process was repeated for aperture diameters from 2 to 40 pixels 
(0\farcs 4--8\arcsec).  We then fit a general power--law function to the 
$\sigma_{\rm aper}$ vs.~$\sqrt{N}$ curve to determine the average
value of $\beta$ over each image; these $\beta$ parameters were
near 1.5 for all images (ranging from $\beta=1.37$ for $J$ to
$\beta=1.58$ for $i^\prime$).

Ideally, the power--law normalization should simply be equal to 
the local pixel--to--pixel RMS $\sigma$.  However, in real images some
deviation from this is possible.  To accurately determine this relation, 
we performed the aforementioned procedure 
over $\sim 300$ random, small ($2000\times 2000$ pixel) subregions of each 
image, this time fitting a function of the form
$\sigma_{\rm aper} = \alpha \sigma (\sqrt{N})^\beta$ and fixing $\beta$
to the average value measured for each mosaic.  The value of $\sigma$
for the pixels contained within the apertures was also calculated, and
a linear fit performed to determine the constant of proportionality
$\alpha$.  For the six images $\alpha$ averages about 1.3 (ranging between 
1.22 and 1.40),
reasonably near the theoretically--expected value of 1.  For each
object in the catalog, the local value of $\sigma$ and the above formula
was used to estimate the background uncertainty, and this was combined
in quadrature with the SExtractor--derived Poisson noise to compute the total flux
uncertainty.  

\begin{deluxetable}{lccc}
\tablecolumns{4}
\tablewidth{200pt}
\tablecaption{UDS/SXDS/SWIRE data characteristics \label{tab_limits}}
\tablehead{
\colhead{Survey\tablenotemark{1}} &
\colhead{Band\tablenotemark{2}} &
\colhead{Area\tablenotemark{3}} &
\colhead{$m_{5\sigma}$\tablenotemark{4}}
\\
\colhead{} &
\colhead{} &
\colhead{(deg$^2$)} &
\colhead{(AB)}
}
\startdata
SXDS &$B$         &0.70   &27.65 \\
SXDS &$R$         &0.70   &27.05 \\
SXDS &$i^\prime$  &0.70   &26.82 \\
SXDS &$z^\prime$  &0.70   &25.53 \\
UDS  &$J$         &0.77   &23.93 \\
UDS  &$K$         &0.77   &23.57 \\
SWIRE&3.6$\mu$m   &0.76   &22.25 \\
SWIRE&4.5$\mu$m   &0.76   &21.53
\enddata
\tablenotetext{1}{SXDS --- Subaru--XMM Deep Survey; UDS--- 
UKIDSS Ultra--Deep Survey; SWIRE---Spitzer Wide-Area Extragalactic Survey.\\}
\tablenotetext{2}{The photometric systems are: Johnson--Cousins ($B$/$R$); SDSS ($i^\prime/z^\prime$); Mauna Kea ($J/K$).\\}
\tablenotetext{3}{Survey areas are in square degrees overlapping the UDS.\\}
\tablenotetext{4}{
Limiting magnitudes ($5\sigma$) are calculated from the average noise properties
of each image using the 1\farcs75 color aperture, and scaled up to
the total object flux using a minimal aperture correction factor (i.e., 
assuming a point source) derived from the growth curves in 
Figure~\ref{fig_gcsn}. 
}
\end{deluxetable}

\subsection{Limiting magnitudes and completeness} \label{sec_complete}
From the noise estimates we calculated approximate $5\sigma$ limiting
magnitudes in the 1\farcs75 color aperture for each band, listed in 
Table~\ref{tab_limits}.  These
are based on the aperture noise measurements in
each image as described in \S\ref{sec_photerr}; the actual limiting magnitudes 
vary as a function of position in each mosaic, particularly
in the SXDS mosaics.  The magnitudes listed in the table have been corrected
for flux falling outside the color aperture (assuming a point source), as
determined from the growth curves shown in Figure~\ref{fig_gcsn}.  Note
that these limiting magnitudes are slightly fainter than those
reported by the survey teams; this is due to our use of a smaller color
aperture (1\farcs 75 vs. 2\arcsec).

\begin{figure}
\plotone{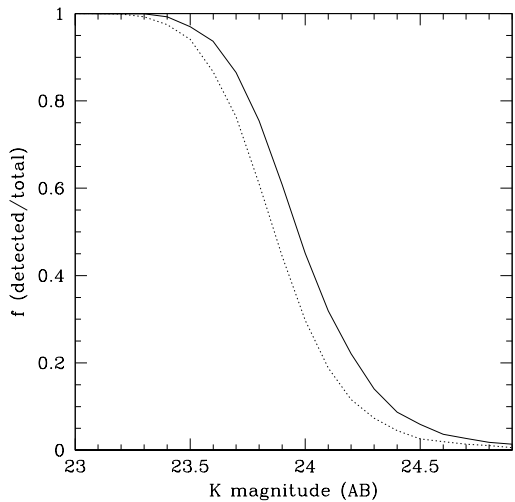}
\caption{Point--source completeness of the $K$--selected catalog, for
the average mosaic depth ({\it solid line}) and the least sensitive
part of the mosaic ({\it dotted line}).
\label{fig_complete}}
\end{figure}

Completeness was tested by placing simulated point sources in a 
$5000\times 5000$ pixel subregion 
of the unconvolved UDS $K$ image.  This subregion sits exactly
in the center of the full mosaic, and was chosen to contain roughly
equal contributions from the four WFCAM arrays (which individually have
slightly different sensitivities), thereby providing a representative sample
of the full mosaic. The $K$ PSF generated for the PSF matching step was
used to create the simulated stars.   Approximately 1700 such stars,
scaled to magnitudes between $K=18-23$, were placed in the image 
on a semi--random grid designed to avoid other real or simulated sources.
SExtractor was then run on the image with the point sources included
(using identical detection parameters to those in \S\ref{sec_sextractor}),
and the number of detected fake stars as a function of magnitude recorded.
At the mean $5\sigma$ $K$--band detection limit of $K=23.57$, the survey
is 95\% complete.

\begin{figure}
\plotone{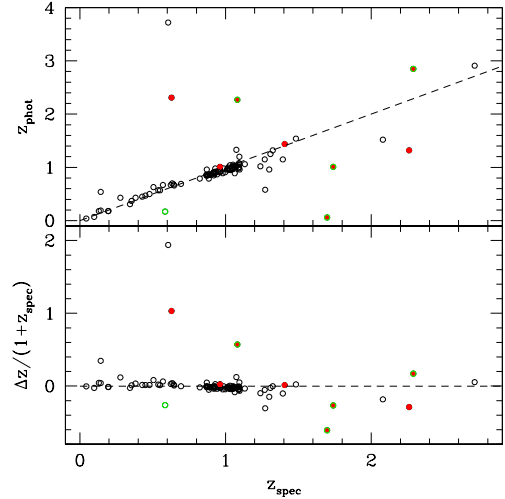}
\caption{Comparison of our photometric redshifts to 119 spectroscopic redshifts in the UDS field ({\it top panel})
and fractional error ({\it bottom panel}).  Red filled circles denote those
points flagged as QSOs in the \citet{simpson06} sample, and green open 
circles are objects with $\log(\chi^2)>2.9$ as derived by EAZY.  
\label{fig_zcompare}}
\end{figure}

This process was repeated for one of the lowest--sensitivity regions
in the mosaic, the area of a single
WFCAM chip in the northwest quadrant of the image.  Even in this worst
case the catalog is 100\% complete for $K<23.2$ and 95\% complete for 
$K\la 23.5$ (see Figure~\ref{fig_complete}).  
It should be noted that these completeness estimates are entirely based
on point sources; for real galaxies with extended flux distributions,
the completeness at a given magnitude will be lower.  However,
for the relatively bright galaxies considered in this analysis 
($K<22.4$, or 1 magnitude brighter than the worst--case 95\% completeness 
limit), this should not be an issue.  Also, 
this does not take into account the possibility of close pairs of objects
being blended, and thus mistakenly being counted as a single object
by SExtractor; however, for this analysis such effects are likely
to be small (further discussed in \S\ref{sec_discussion}).

\subsection{Catalog format}
The final $K$--selected catalog employed in this analysis was generated from 
the 99022 objects (not lying near bad pixels) detected by SExtractor in the 
UDS mosaic.  Some of these may be image artifacts and/or fall on bad
or missing regions in the SXDS and SWIRE images as described in 
\S\ref{sec_badpix}.  Such objects are 
included in the catalog, but with flag(s) noting that their photometry in a
given band is likely to be unreliable.
Photometry in the $B$ through IRAC 4.5\,$\mu$m bands, as well
as limited morphological information, are included, and all fluxes
are given with an AB zeropoint of 25.  The columns of
the catalog\footnote{Available from\\
\url{http://www.strw.leidenuniv.nl/galaxyevolution/UDS}} are as follows:

\noindent
{\it Columns (1)--(3).}---Running ID, right ascension, and declination (J2000)\\
{\it Columns (4)--(15).}---$\brizjk$ ``color'' fluxes and errors (1\farcs 75 
diameter aperture, listed in the order $f_B$, $\sigma_B$, $f_R$, $\sigma_R$, 
...)\\
{\it Column (16).}---Total $K$ flux in the SExtractor {\tt AUTO} aperture\\
{\it Columns (17)--(20).}---IRAC 3.6$\mu$m and 4.5$\mu$m fluxes and errors 
(matched to the optical/IR ``color'' aperture)\\
{\it Columns (21)--(23).}---$K$--band half--light radius, ellipticity, and position angle\\
{\it Columns (24)--(25).}---Optical and SWIRE bad--pixel flags\\
{\it Column (26).}---Flag for objects not detected in the stacked 
$B+R+i^\prime+z^\prime$ image\\
{\it Column (27).}---Internal flag generated by SExtractor

\section{Photometric Redshifts and rest--frame colors} \label{sec_photz}
\subsection{EAZY fitting} 
Photometric redshifts were determined with the EAZY code\footnote{Code
and documentation are available at \url{http://www.astro.yale.edu/eazy}}, 
described in
detail in \citet{brammer08}.  In its default configuration, EAZY uses
$\chi^2$ minimization to fit linear combinations of six basis templates to 
broadband galaxy spectral energy distributions; a $K-$band luminosity
prior and estimates of systematic errors due to template mismatch are
also taken into account.  These default settings 
have been demonstrated to provide reliable photometric redshifts for
other $K$--selected samples \citep[see][]{brammer08}.

To test the reliability of the UDS $\zphot$ values we compared them
to galaxies in this field with known spectroscopic redshifts.  A query using 
the {\it NASA Extragalactic Database} (NED) finds 96 published redshifts, 
most of which are old passively--evolving galaxies at $0.8<z<1.2$ from
\citet{yamada05}.  An additional 60 redshifts of radio sources
are from the catalog of \citet{simpson06}.  After 
cross--correlating these objects with objects detected in the UDS $K$ band, 
119 spectroscopic redshifts remained; EAZY was able to find $\zphot$ 
solutions for 110 of these (and found no $\chi^2$ minimum for the other 9).  
The left--hand panel of Figure~\ref{fig_zcompare} shows a comparison
of the photometric and spectroscopic redshifts of these galaxies.
Objects flagged by \citet{simpson06} as ``QSO'' or ``XQSO''
(X-ray emitting QSO), which may not be well--fit
by typical galaxy templates if the AGN light contributes significantly
to the overall flux, are shown as filled red circles in the plot.  Of the
aforementioned nine objects for which EAZY failed to find $\zphot$ solutions, 
four of them fell in this AGN category.

While the agreement between the photometric and spectroscopic redshifts
is quite good (particularly at $z\sim 1$), roughly 8\% of the points are 
outliers with fractional errors of 0.2 or more (and about half of these in turn
are serious outliers,
with $\left|\Delta z\right|/(1+z_{\rm spec})>0.5$).  Determining {\it a priori}
from photometry alone which objects are likely to be outliers is 
difficult -- although most of the QSOs are indeed strong outliers, weak
(or even strong) AGN activity in galaxies is not always evident.  However,
the $\chi^2$ value returned by EAZY appears to be a somewhat reliable
indicator: all objects with $\log \chi^2>2.9$ 
have $(1+\zphot)$ values that deviate more than 20\% from their
measured spectroscopic redshifts.  When objects with these high $\chi^2$
values are removed from the sample, 
the remaining normalized median absolute deviation (NMAD) in 
$\Delta z/(1+z_{\rm spec})$ is 0.033 with a median offset
of $\Delta z=-0.013$.  

\begin{figure}
\plotone{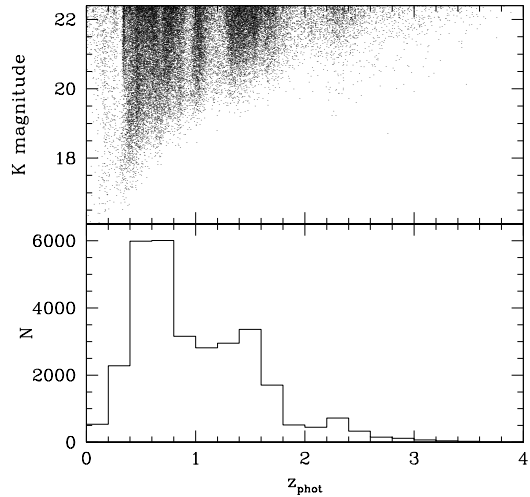}
\caption{Distribution of apparent $K$ magnitude as a function of photometric
redshift ({\it top panel}), and number of galaxies as a function of
$\zphot$ ({\it bottom panel}).  
\label{fig_zdist}}
\end{figure}

At $z\sim 1.2$ the photometric redshifts appear to exhibit a small
systematic offset of $\Delta z/(1+z)\sim -0.05$ (though due to the small 
number of points here it
is difficult to assess the magnitude of the offset).  A similar offset at 
the same redshift is seen by \citet{brammer08} even with more
photometric bands included (notably $U$, $V$, and $H$), and the offset 
persists when different template sets and input parameters are
used.  Additionally, relatively small--scale ``spikes'' and ``voids''
(on the scale of $\Delta z=0.1-0.2$) in the photometric redshift 
distribution are apparent--see Figure~\ref{fig_zdist}.  These may 
in part represent real variations in the large--scale galaxy
distribution, but some are also likely to be numerical artifacts intrinsic to
the photometric redshifts.  However, since these effects are relatively
small, and since in our analysis we employ redshift intervals
($\Delta z>0.5$) larger than the observed ``spikes'' in the $\zphot$ 
distribution, such effects are not likely to significantly affect our
results.

The rest--frame near-- and mid--infrared properties of typical galaxies 
are not as well constrained as the optical SEDs, so the
galaxy templates employed by EAZY tend to be uncertain in this regime.
This uncertainty
is taken into account during the $\zphot$ calculations; nevertheless, 
the inclusion of the IRAC data could still in principle introduce
systematic effects.  To check this, we re--ran
EAZY with only the $\brizjk$ photometry and compared the
results to those with the IRAC fluxes included.  While the resulting
$\zphot$ values were similar, the scatter and systematic offset increased
substantially, with ${\rm NMAD}\left[\Delta z/(1+z_{\rm spec})\right]=0.044$ 
and median $\Delta z=-0.017$.  Moreover, although the number of major outliers
is comparable, such outliers are no longer easily rejected: even with 
the $\chi^2$ cut and excluding objects where solutions were not found,
only three of the AGNs are excluded, and overall the worst outliers
are no longer confined to the fits with the highest $\chi^2$ values.

Including IRAC data in the calculation of photometric redshifts
thus reduces the $\zphot$ scatter and allows outlying points (including
many AGNs) to be rejected from the sample .  This is
perhaps not surprising: for
example, \citet{stern05} find that AGNs typically have redder
IRAC $[3.6]-[4.5]$ colors than typical galaxies.  Since the galaxy templates
employed by EAZY do not take possible AGN components into account, the 
4.5\,$\mu$m excess
in galaxies with nuclear activity in turn cannot be matched well by any
of these templates.  Indeed, most of the
objects that either had high $\chi^2$ values or no solutions from EAZY
appeared to exhibit redder $[3.6]-[4.5]$ colors than those galaxies with
good $\zphot$ fits.

\begin{figure}
\plotone{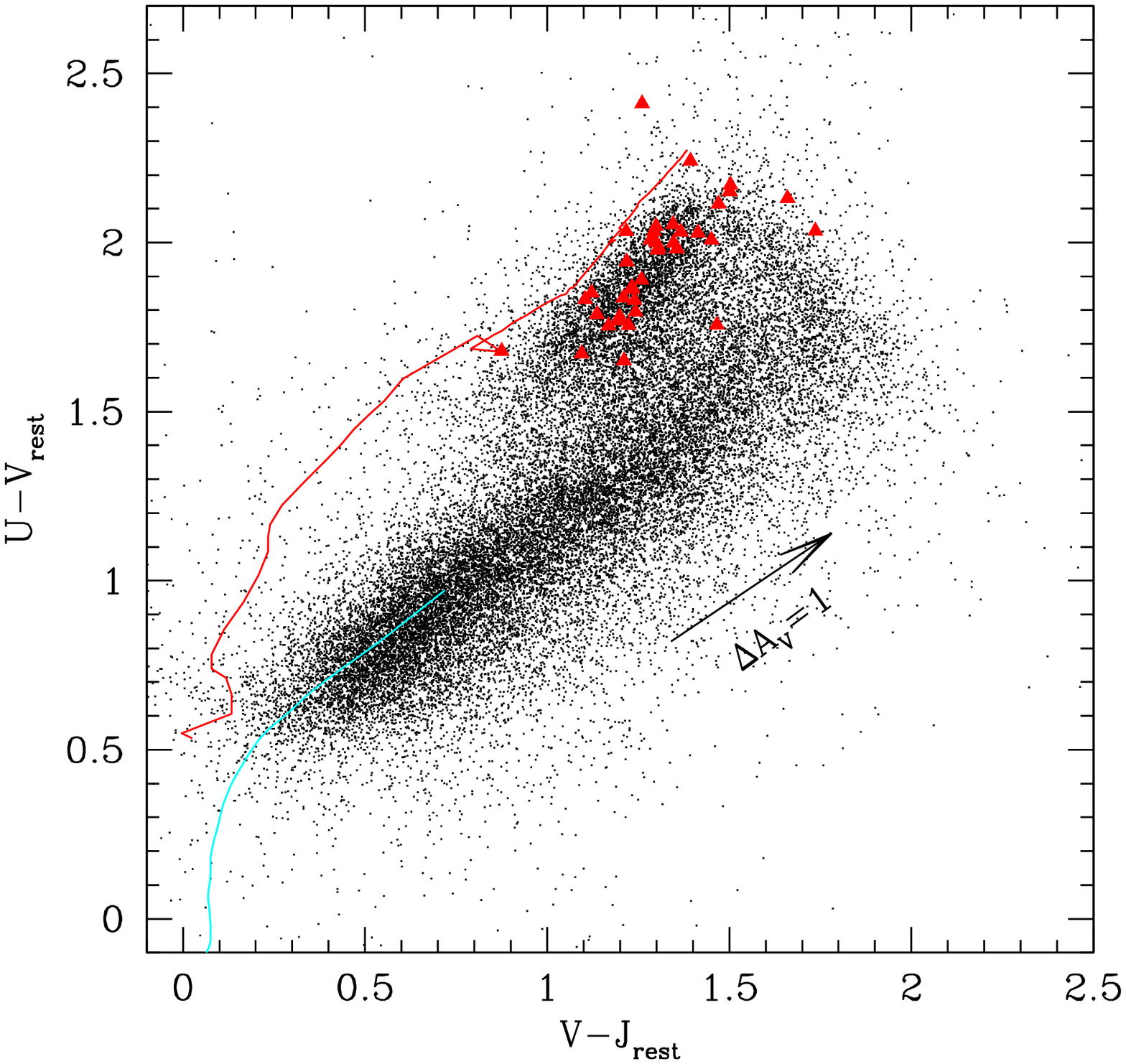}
\caption{Rest--frame color--color diagram for $K_{\rm AB} <22.4$, $\zphot<2.5$
UDS galaxies ({\it black points}).  Red points denote
spectroscopically--confirmed $z\sim 1$ old passively--evolving galaxies (OPEGs)
from \citet{yamada05} with little or no detected line emission 
($W_\lambda({\rm OII})<5$\,\AA).  \citet{bc03} evolutionary tracks of
passively--evolving ({\it red line}) and constantly star--forming ({\it cyan
line}) stellar populations from $0.1-10$\,Gyr are overplotted,
and the arrow shows the effect of 1 magnitude of
dust extinction.  The extended star--forming track
and the ``quiescent clump'' (which overlaps with the OPEG sample) are
clearly separated.
\label{fig_rfcol}}
\end{figure}

\subsection{Monte Carlo analysis} \label{sec_montecarlo}
During the process of computing $\zphot$ values, EAZY also provides
estimates on the redshift uncertainty and a probability distribution
$p(z)$ for each object.  The $p(z)$ curves in particular are useful
for estimating the redshift distribution of a large sample of objects,
since the true distribution is likely to be broader than a histogram
of the best--fit $\zphot$ values (e.g., in an extreme case with a sample
of $\zphot=2.00$ galaxies, the true distribution would be much broader
than a delta function).  To assess the consistency of the $p(z)$ distributions,
we performed 120 Monte Carlo iterations
varying the input photometry assuming Gaussian errors.  Though this
is a relatively small number of iterations, it gives a rough
estimate of the $\zphot$ uncertainty for any given object, and an accurate
computation of the redshift distribution of a large sample of objects.

For each object, the median and dispersion among the 120 Monte Carlo
runs were determined and compared to the estimates from EAZY.  The median
values for the UDS sample were consistent with the best--fit values
derived by EAZY from the original (unperturbed) catalog, as were the
$1\sigma$ uncertainty estimates.  Furthermore, for subsets of the
catalog, the redshift distributions derived from the Monte Carlo analysis
closely reflected the EAZY--derived $p(z)$ distributions.  The uncertainties
derived by EAZY thus appear to be internally consistent for the UDS data.

\begin{figure*}
\plotone{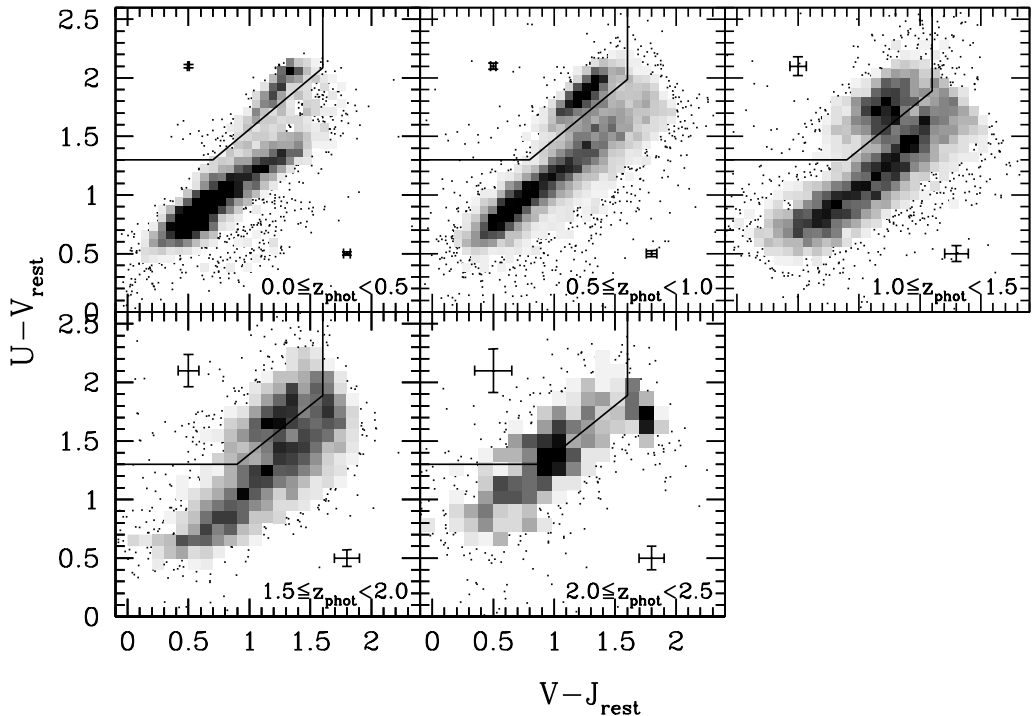}
\caption{Rest--frame $U-V$ vs. $V-J$ colors of UDS $K<22.4$ galaxies
in five redshift bins.  The greyscale represents the density of points
in the central region of each plot, while outlying points are plotted
individually.  
The solid lines show the adopted divisions between the
star--forming and quiescent galaxy samples at each redshift, defined
in Equation~\ref{eq_division}.  Median 
uncertainties in the rest--frame colors of quiescent and star--forming galaxies
(only taking random errors into account) are shown in the 
upper left and lower right of each frame, respectively. 
\label{fig_evol}}
\end{figure*}

\subsection{Interpolating rest--frame photometry}
At $z\sim 2.5$, the $U$, $V$, and $J$ rest--frame bands fall roughly into
the observed $J$, $K$, and IRAC $4.5\mu$m bands 
respectively.  Up to this redshift, intrinsic $UVJ$ fluxes can thus be 
interpolated from the observed data.  Filter response curves for the observed 
bands, taking into account atmospheric
absorption and detector quantum efficiency, were downloaded from each of the
WFCAM, Suprime--Cam, and IRAC websites.  For rest--frame filter
definitions we used the standard filter definitions without atmospheric
absorption or detector response included: the \citet{bessell90} $U$ and $V$ 
curves, and the Mauna Kea definition \citep{tokunaga02} for $J$.
Rest--frame fluxes were then interpolated following the 
method of \citet{rudnick03}\footnote{This was
carried out using the InterRest script (E. Taylor et al., in preparation);
see \url{http://www.strw.leidenuniv.nl/~ent/InterRest}}.
Uncertainties on the rest--frame fluxes were derived using the perturbed input
catalogs and output redshifts from the Monte Carlo analysis described
in \S\ref{sec_montecarlo}.  Note that these color uncertainties only take
into account photometric errors (and the resulting $\zphot$ errors),
and do not include the uncertainties intrinsic to the templates used for 
$\zphot$ fitting.

\section{Evidence for quiescent galaxies to $z\sim 2.5$} \label{sec_analysis}
\subsection{The rest--frame colors of quiescent and star--forming galaxies}
We use the galaxy catalog derived from the UDS/SXDS/SWIRE data,
along with the photometric redshifts and interpolated rest--frame
colors described in the previous section, to analyze the 
rest--frame color distribution of galaxies out to $z=2.5$.  Beyond this
redshift the rest--frame $J$ band begins to ``fall off'' the reddest
filter in our catalog (IRAC $4.5\mu$m), and rest--frame colors
at $z>2.5$ are therefore less reliable.  
Stars are selected (and removed from the sample) using the two criteria:
$J-K<0$ and $(J-K)<0.2(i^\prime-K)-0.16$.  This two--color cut was
derived by inspection of the $J-K$ vs.~$i^\prime-K$ color--color diagram,
wherein stars form a tight, well--defined track.  Additionally, we exclude 
all objects that fall on bad pixels or missing data regions in any band,
as well as those with bad photometric redshift solutions ($\log\chi^2>2.9$
as derived in \S\ref{sec_photz}), and apply a magnitude limit of
$K<22.4$.  When all these criteria are met, the subsample analyzed
hereafter contains 30108 galaxies between $0<z<2.5$.  By comparison
with 119 spectroscopic redshifts in the field we find typical
photometric redshift errors of $\Delta z/(1+z)\sim 0.033$, though
this is almost entirely measured with galaxies at $z\la 1.2$; at higher 
redshifts the $\zphot$ uncertainty is likely to be substantially larger.

Figure~\ref{fig_rfcol} shows the rest--frame $U-V$ vs. $V-J$ (hereafter
$UVJ$) diagram for this subset of UDS galaxies.  A striking bimodality
emerges: one diagonal track extends from
blue to red $V-J$, while a localized clump that is red in $U-V$ but
blue in $V-J$ lies above this track.  Previously,
\citet{labbe05} and \citet{wuyts07} found that actively star--forming
and quiescent galaxies segregate themselves in this plane, with
the star--forming galaxies forming a diagonal track and
quiescent galaxies populating mostly the upper left--hand region; with the
far larger number of sources in the UDS, it is now evident that the galaxies
form a truly bimodal distribution in this plane.  Indeed,
this interpretation is supported here by both data and models:
the red points overlying the ``quiescent clump'' in Figure~\ref{fig_rfcol} 
are spectroscopically--confirmed old passively--evolving galaxies 
from \citet{yamada05}, while the star--forming and quiescent loci are 
reasonably coincident with the corresponding \citet{bc03} stellar population 
models; a detailed analysis of the star--formation properties of galaxies
in the $UVJ$ plane will be presented in a forthcoming paper (Labb\'e et al.,
in preparation).

\begin{figure*}
\plottwo{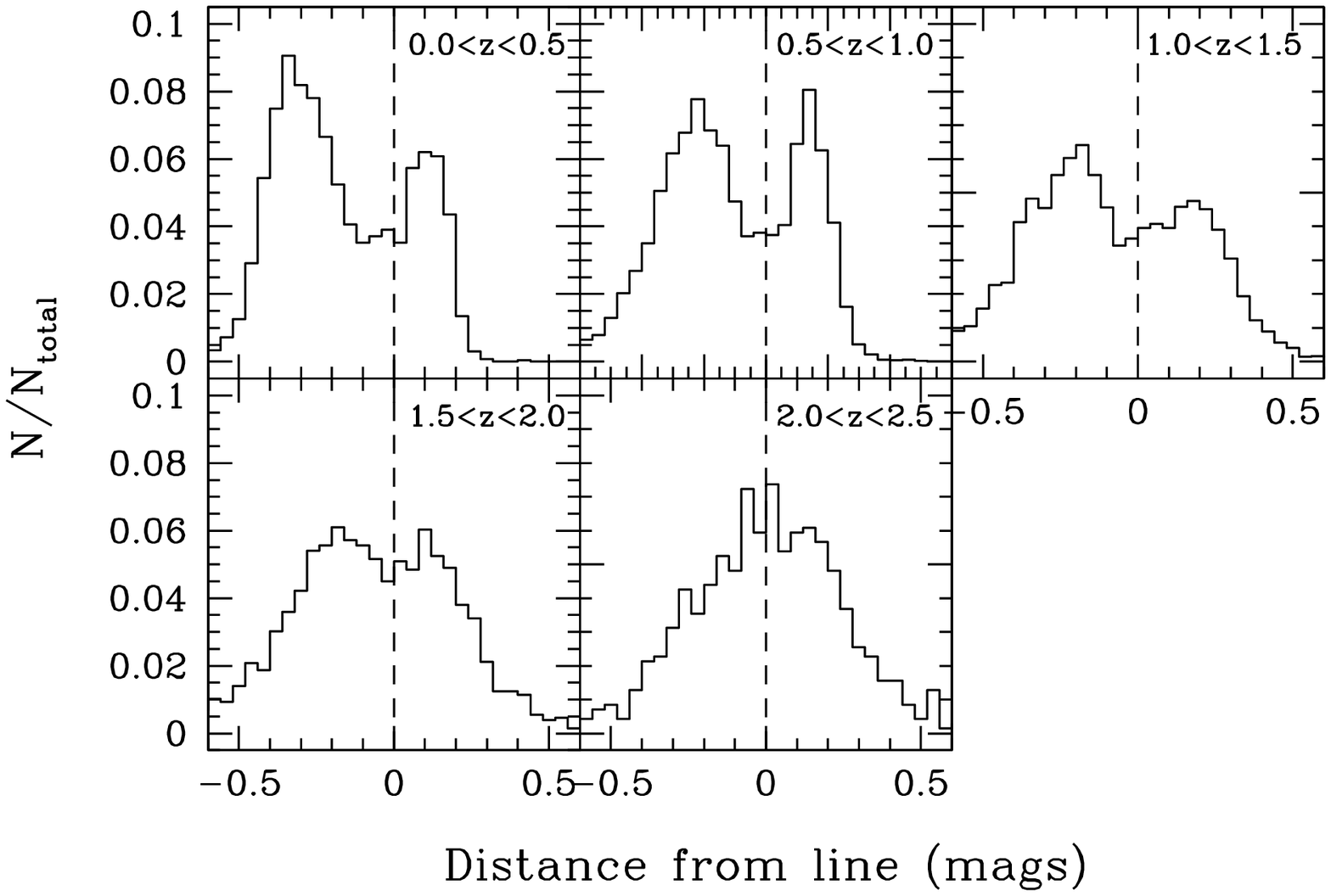}{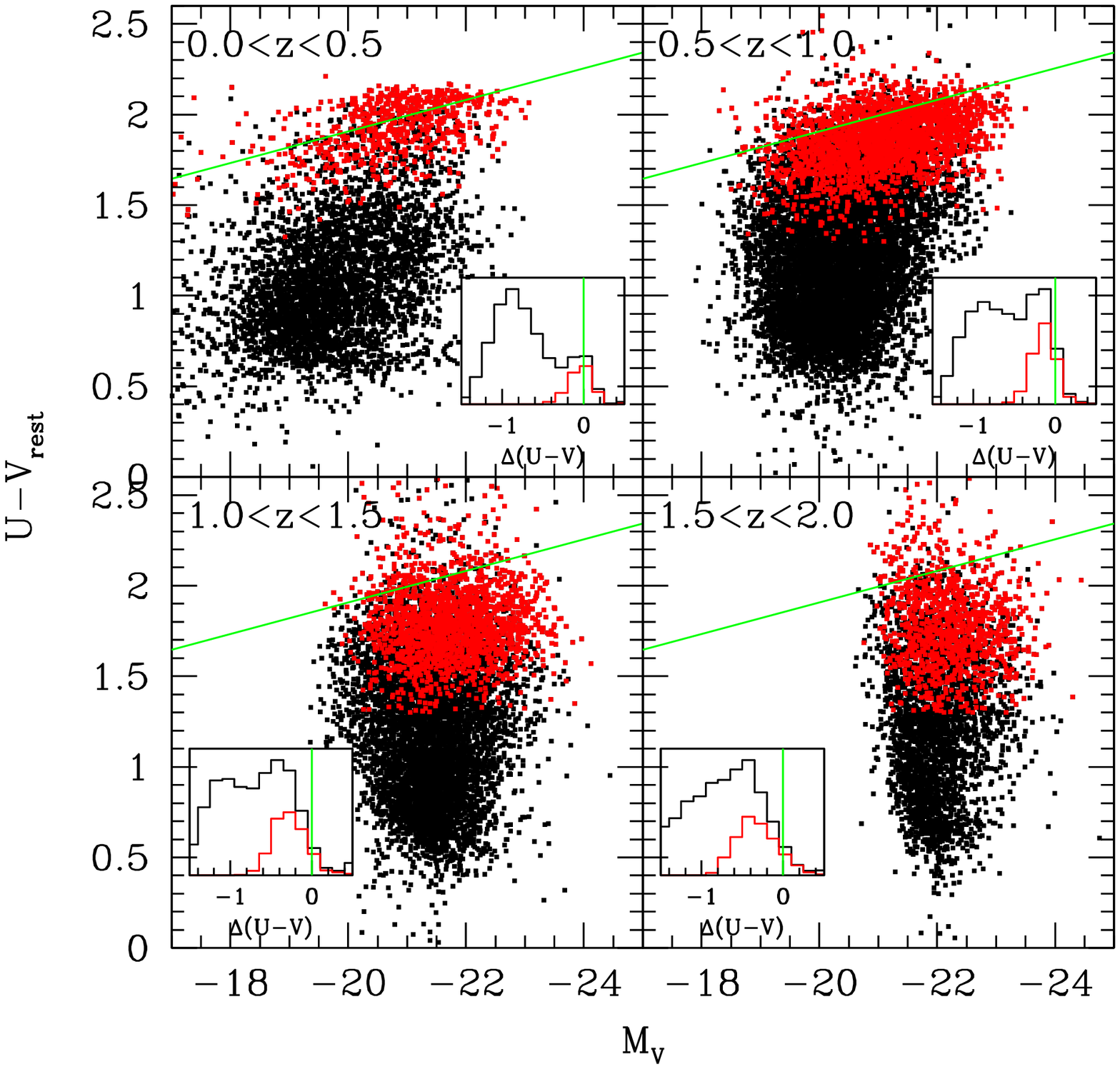}
\caption{{\it Left panel:} Number of red galaxies 
as a function of the normal distance from the diagonal dividing lines shown in 
Figure~\ref{fig_evol}; the normal distance is defined such that,
in these histograms, star--forming galaxies are to the left (i.e.
negative values) and quiescent galaxies are on the right.
Bimodality along this diagonal line is
clearly evident up to $z=2$.  {\it Right panel:} Color--magnitude
diagrams in the same bins; red and black points denote quiescent and
star--forming galaxies (selected on the basis of the cuts in 
Figure~\ref{fig_evol})
respectively.  The green line is a fit to the red points in the $0.0<z<0.5$
bin, and the inset plots show galaxy counts as a function of the distance
in $U-V$ from this line (with quiescent galaxies plotted separately
as a red histogram).  Note that no $U-V$ bimodality is evident at $z>1.5$,
even though it is clearly seen in the $UVJ$ plane.
\label{fig_dahist}}
\end{figure*}

Essentially, the $UVJ$ diagram allows the degeneracy between
red star--forming and red quiescent galaxies to be broken: 
while galaxies with blue $U-V$ colors
in general exhibit relatively unobscured star formation activity, red 
galaxies could be either
quiescent galaxies with evolved stellar populations or dust--obscured
starbursts.  But since dust--free quiescent galaxies are {\it blue} in
$V-J$, they occupy a locus in the $UVJ$ plane that is distinct from
the star--forming galaxies, allowing the two populations to be empirically
separated.  Clearly, such a separation using a single color (such as
$U-V$) would be fraught with problems: at best, the quiescent sample
derived in this manner would be contaminated by red starbursts, and if
the number of such starbursts were sufficiently high the bimodality would 
no longer be visible (see also Figure~\ref{fig_dahist}).  Note that
although the $U-B$
color is better at distinguishing between a narrow break (characteristic of
an old stellar population) and broader dust reddening when spectroscopic
redshifts are available, the larger uncertainties on photometric redshifts
do not allow sufficiently accurate $U-B$ colors to be estimated.

Even when photometric 
redshifts are derived using alternative template sets within EAZY, 
or an entirely different $\zphot$ code \citep[HYPERZ;][]{hyperz}, 
the basic bimodal shape of Figure~\ref{fig_rfcol}
persists, indicating that this method is robust to the specific
numerical technique employed.

\subsection{Color Evolution}
Changes in rest--frame colors with redshift reflect the intrinsic evolution of 
stellar populations, since all spectra have in 
principle been transformed to the same reference frame (barring possible 
$\zphot$ systematic errors).  $UVJ$ diagrams at five different 
redshifts (in bins of width $\Delta \zphot=0.5$) are shown in 
Figure~\ref{fig_evol}.  The most immediately apparent feature is that
the observed bimodal distribution of star--forming and quiescent galaxies 
is clearly seen up to $z\sim 2$. Although the scatter increases substantially 
at higher redshifts (to the point where it likely washes out any intrinsic
bimodality at $z>2$), most likely due to a combination of larger 
photometric redshift uncertainties and weaker observed fluxes, the two
distinct populations are nonetheless visible.  

It is also notable that the shape
of the star--forming sequence appears to change: at the lowest redshifts
the sequence curves, but this may be due to the effects of using a small
aperture on relatively nearby galaxies (i.e.,
the outermost, bluer parts of galaxies with large angular sizes falling
outside the ``color'' aperture).  
At $z\ga 0.5$, where the angular size--redshift relation begins to flatten,
the dust sequence indeed becomes more linear.
The increase in star formation activity at higher redshifts is also
apparent in two ways: the fractional number of dusty, star--forming
galaxies (the upper--right portion of the dust sequence) increases,
and the entire dust sequence appears to move to redder $V-J$ colors.  
Note that at $2.0<\zphot<2.5$ a concentration at red $V-J$ and $U-V$ appears;
this mainly comprises galaxies with a substantial contribution from
the red $A_V=2.75$ template included with EAZY.  This single template
substantially improves the $\zphot$ solutions of very dusty galaxies, but 
can lead to the observed discrete clump in color--color space.

With increasing redshift the dead clump appears to steadily move to 
bluer $U-V$ (with the median $U-V$ changing by $\sim 0.15$ magnitudes
between the $z=0-0.5$ and $z=1.5-2$ bins), while the
$V-J$ color remains more or less unchanged.  This is expected from
passive evolution of the stellar populations in the clump, but
relatively small systematic errors in the photometric redshifts 
(on the order of $\Delta z/(1+z)\sim 0.05$) can produce similar offsets
in the average $U-V$ color of quiescent galaxies at $z\sim 1.5$.
The $\zphot$ estimates show a
systematic offset that is smaller than this, about 
$(\zphot-z_{\rm spec})/(1+z_{\rm spec})\sim -0.01$ over the full redshift
range (see Figure~\ref{fig_zcompare}).
Nonetheless, given the lack of spectroscopic redshifts at $z\ga 1.3$, we
cannot ascertain whether there are larger systematic offsets at higher
redshift, and therefore cannot accurately measure the color evolution 
of galaxies in the quiescent clump or absolutely define the boundaries of the
box in which they live.

\subsection{Sample Division}
Since the color bimodality is visible up to $z\sim 2$, we employ
empirical criteria to divide this sample into quiescent and
star--forming subsamples.  In each redshift bin of Figure~\ref{fig_evol} 
where the bimodality could be seen,
an initial diagonal cut was first made between
the two populations.  Histograms of red galaxy counts relative to this
line were then derived.  The position of the 
diagonal cut was fine--tuned (keeping the same slope at all redshifts) 
to fall roughly between the two peaks (except in the lowest--redshift bin,
where the shapes of the quiescent and star--forming galaxy loci made
this impossible); Figure~\ref{fig_dahist} (left panel) shows these histograms 
along the adjusted diagonal lines. 

The adopted diagonal selection criteria for quiescent galaxies between
$0<z<2.0$ bin are:
\begin{equation}
\label{eq_division}
\begin{array}{lcccr}
(U-V)>0.88\times(V-J)+0.69 & & & &\left[0.0<z<0.5\right]\\
(U-V)>0.88\times(V-J)+0.59 & & & &\left[0.5<z<1.0\right]\\
(U-V)>0.88\times(V-J)+0.49 & & & &\left[1.0<z<2.0\right]
\end{array}
\end{equation}
Additional criteria of $U-V>1.3$ and $V-J<1.6$ are applied to the
quiescent galaxies at all redshifts to prevent contamination from 
unobscured and dusty star--forming galaxies, respectively.  
The samples of star--forming galaxies are then defined by everything 
falling outside this box (but within the color range plotted in
Figure~\ref{fig_evol}, such that the very small number of extreme color 
outliers are not included in either sample).
The two distinct populations are no longer visible at $\zphot>2.0$,
but the same dividing line used at $z=1-2$ is shown for reference.  
Note that the exact placement of this division does not significantly
affect the results presented herein (see also \S 7 for a more detailed
discussion).
Typical random color uncertainties for the star--forming and quiescent
samples at each redshift interval are shown in the lower right and upper
left corners, respectively, of each panel in Figure~\ref{fig_evol}.

\begin{figure}
\plotone{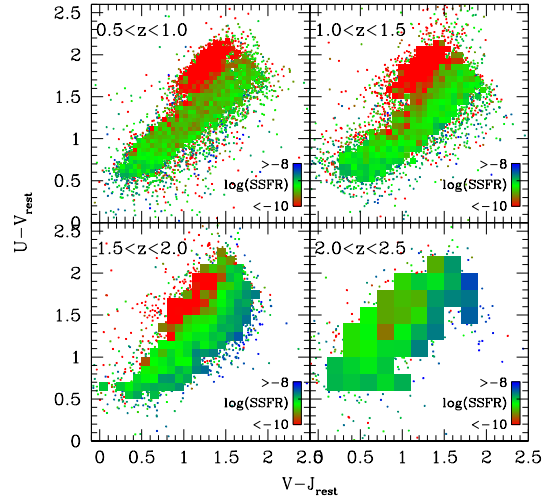}
\caption{Rest frame $UVJ$ diagram in four redshift ranges
with bins color coded by the median specific star--formation rate of 
the galaxies within each bin; outliers in the $UVJ$ plane are plotted
individually.  The
$z<0.5$ bin is not included here because the adopted SSFR proxies are not 
reliable at low redshifts.  Quiescent galaxies lie in approximately 
the same region at all redshifts.
\label{fig_div24um}}
\end{figure}

\subsection{MIPS $24\mu$m measurements} \label{sec_mips}
Between $z\sim 1-2.5$, the $24\mu$m flux of galaxies is strongly correlated
with dust--obscured star formation due to the presence of redshifted
$6-12\mu$m polycyclic aromatic hydrocarbon (PAH) features in this
band \citep[e.g.]{yan04}.  The observed $24\mu$m fluxes of the UDS galaxies
can thus be used to empirically confirm whether the $UVJ$ criteria indeed
select star--forming and quiescent galaxies.  
Since the $K$ band, which at these redshifts corresponds to rest--frame
optical/NIR light, is a rough tracer of stellar mass, the $24\mu$m/$K$ flux
ratio then provides an estimate of the dust--obscured specific 
star--formation rate (SSFR).  Likewise, the UV--to--$K$ flux ratio provides
an approximate measure of the unobscured SSFR, and thus the combination of
these three bands provides a reasonably robust estimate of total
SSFR.
  
Using the {\it FIREWORKS}
catalog of \citet{wuyts08} we derive proxies
for the IR and UV SSFR based on UV/$K$ and $24\mu m/K$ flux ratios, and
use these proxies to obtain order--of--magnitude estimates of the SSFRs
of UDS galaxies; the details of this derivation are given in the Appendix.
Figure~\ref{fig_div24um} shows the
$UVJ$ diagram in four redshift bins between $0.5<z<2.5$ (the 
bin at $z<0.5$ is not shown, since the SSFR proxies are not expected to
be effective at such low redshifts).  
Since the $24\mu$m data are too 
shallow for most galaxies in this sample to be detected individually,
we combine the SSFRs of multiple galaxies as a function of $UVJ$ color as
follows:  at each redshift
galaxies are divided into square $UVJ$ color bins. If $>10$ objects fall into
a given bin, a square is plotted with a color corresponding to
the {\it median} SSFR of the underlying galaxies
(in essence ``binning'' the galaxies' SSFR estimates); otherwise, individual 
points are plotted.
A strong trend is clearly seen from $z=0.5-2$: the ``quiescent 
clump'' on average exhibits low SSFR, and is distinctly offset from the
adjacent red star--forming galaxies.

\begin{figure}
\plotone{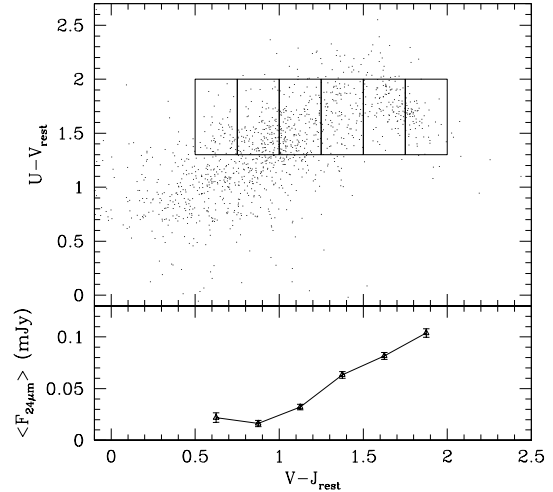}
\caption{Rest--frame colors of bright ($K<22.4$), $2.0<z<2.5$ galaxies
({\it top panel}).  Red galaxies with $1.3<(U-V)<2.0$ are selected
by the boxes shown, and the median $24\mu$m MIPS flux for the galaxies 
within each box is plotted in the lower panel.  The $24\mu$m flux rises at
red $V-J$, confirming the trend suggested in the corresponding panel of
Figure~\ref{fig_div24um}.
\label{fig_rf24um}}
\end{figure}

In the $2.0<z<2.5$ bin of Figure~\ref{fig_evol} the scatter in the $UVJ$
diagram is too large for any intrinsic bimodality to be seen,
and the trend in SSFR also appears weaker (Figure~\ref{fig_div24um}), 
perhaps due in part to the 
smaller number of points at high redshift and passive evolution.  At 
$z>2$, we employ another approach to verify
the presence of galaxies with relatively low SFRs: the top panel of
Figure~\ref{fig_rf24um}
again shows the $z=2-2.5$ $UVJ$ plot, with red galaxies selected by
the rectangular boxes shown.  The median $24\mu$m flux of the $N$ galaxies
within each box, measured in 8\arcsec\ diameter apertures, is then 
calculated.  To estimate the uncertainties in the median
fluxes, $N$ apertures were placed randomly on the MIPS image
and the median $24\mu$m flux within these apertures calculated.  This
was repeated 500 times for each value of $N$, and the clipped standard
deviation of these 500 trials was taken as the uncertainty in the median
of the galaxy fluxes for every bin of $N$ galaxies.  Note that since the 
randomly--placed apertures
may contain bright and/or confusing $24\mu$m sources, this error estimate
inherently takes both background and confusion noise into account.

The median $24\mu$m fluxes within each bin are shown in the lower panel
of Figure~\ref{fig_rf24um}; indeed, $24\mu$m flux increases strongly
with $V-J$ color, indicating that $V-J$ does indeed provide a good
proxy for star formation rate at $z=2-2.5$ (with a factor of $\sim 5$
difference in $24\mu$m flux between the bluest and reddest galaxies
in $V-J$). 
The trends in $24\mu$m flux and SSFR strongly suggest that the $UVJ$
selection criteria indeed separate quiescent from
star--forming galaxies (even given the limited number of photometric
bands available in UDS) at $z<2$, and may work reasonably well
up to higher redshifts, $z\la 2.5$. Since the color bimodality can no longer
be seen at $z>2$ it is possible that the quiescent galaxy sample at these
redshifts contains somewhat more contamination from red, star--forming 
galaxies; furthermore, the rest--frame $J$ flux estimate relies on the 
relatively shallow $3.6/4.5\mu$m data at $z>2$, which may result in subtle 
incompleteness effects in the sample at these redshifts.  Therefore, for 
the remainder of our analysis we focus only
on galaxies at $z<2$ where the bimodality can obviously be seen.  
It is also important to note that the presence of AGN 
may produce enhanced $24\mu$m emission and redder $V-J$ colors, and
thus mimic the photometric effects of star formation.  As described 
in \S\ref{sec_photz}, most objects with known strong AGN activity do
not have valid $\zphot$ solutions and have likely been removed from the
sample; even so, some weak AGN activity may remain and contribute to the
observed correlation.

\begin{figure}
\plotone{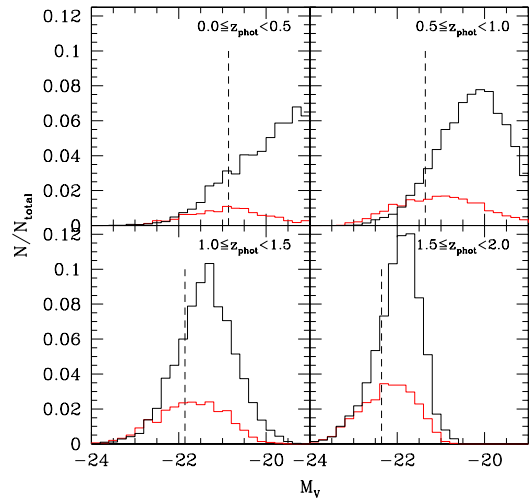}
\caption{Rest--frame $V$ magnitude distributions of the quiescent
({\it red 
histogram}) and star--forming ({\it black histogram}) galaxies in the four
redshift ranges shown in Figure~\ref{fig_dahist}.  The vertical dashed line
shows the value of $M_V^\star$ at the central redshift of each bin,
assuming the $z=0.1$ SDSS value of \citet{blanton03} and evolving
as $M_V^\star(z)=M_V^\star(0)-z$.  At high redshifts,
quiescent galaxies compose roughly half of the total bright galaxy population,
and a larger fraction at lower redshifts.
\label{fig_maghist}}
\end{figure}

\subsection{Luminosities}
Absolute rest--frame $V$ magnitude histograms of the number counts of 
quiescent and star--forming 
galaxies in each redshift interval are shown in Figure~\ref{fig_maghist}.  
The value of $M_V^\star$ at the center of each redshift bin, assuming
the \citet{blanton03} SDSS value at $z=0.1$ and evolving as 
$M_V^\star(z)=M_V^\star(0)-z$, is also shown.
At bright magnitudes ($M_V\la -22.5$)
between $z\sim 1-2$, quiescent galaxies contribute approximately as much 
to the galaxy number density as those that are actively star--forming.
This is similar to the result of \citet{kriek06}, who found star formation
rates consistent with zero in  9/20 bright galaxies at $z=2.0-2.7$, though
their spectroscopic determination of SFRs is likely to produce fundamentally
different ``quiescent'' samples than our rest--frame color definition.
At lower redshifts ($0.5<z<1.0$), the luminous
galaxy population is comprised of an larger fraction of
quiescent galaxies, while galaxies with fainter $V$ luminosities (and hence 
which typically have lower stellar masses) are dominated by those 
undergoing star formation.

\section{Clustering of quiescent and star--forming galaxies} \label{sec_clustering}
\begin{figure}
\plotone{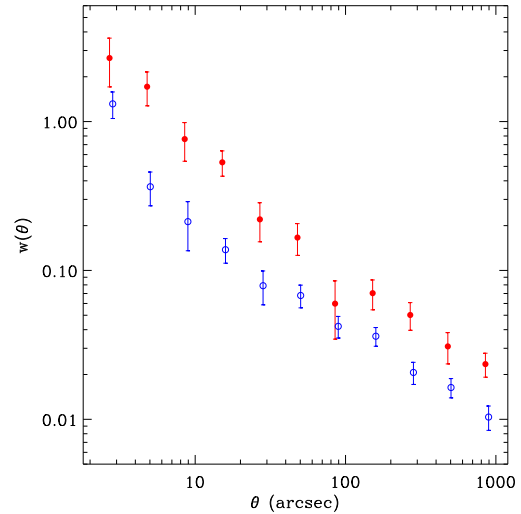}
\caption{Angular correlation functions for the $1.0<\zphot<2.0$ quiescent
({\it red filled circles and errorbars}) and star--forming 
({\it blue open circles }) samples.  The
quiescent galaxy sample shows significantly stronger angular clustering.
\label{fig_acf}}
\end{figure}

To investigate the relation between halo mass and star formation
activity at $z>1$, we calculate the clustering
of the star--forming and quiescent galaxies (as defined by their rest--frame 
$UVJ$ colors) at these redshifts following the method of 
\citet{quadri07a,quadri08}; a brief summary of this method follows.
The angular correlation function of each sample was computed using
the \citet{landy93} estimator, taking into
account all ``bad'' data regions (described in detail in \S\ref{sec_badpix}). 
Figure~\ref{fig_acf} shows these functions for the quiescent
and star--forming samples, with the former clearly exhibiting stronger
clustering (in angular space) than the latter.

Because of the large number of objects in this catalog at $z=1-2$, it
is possible to further break each of the quiescent and star--forming
samples into bright and faint subsamples, thereby also investigating
the effect of luminosity on clustering strength.
Angular correlation functions were recomputed for star--forming
and quiescent galaxies within two luminosity bins divided at 
$M_V=-22.0$.  This is approximately the minimum luminosity at which 
the $z=1-2$ sample is complete---essentially all galaxies with
$M_V<-22$ at $z=2$ have observed $K$ fluxes brighter than the $K<22.4$ cutoff,
while at $M_V>-22$ some galaxies (most of which are low--mass and blue)
fall below the $K$ flux limit.
However, this should not affect the reliability of the 
$r_0$ determination itself since the Monte Carlo $\zphot$ simulations
(\S\ref{sec_montecarlo}) fully account for differences
in the redshift distributions.  For each subsample we fit power laws to  
the angular correlation functions over the range $60\arcsec - 300\arcsec$,
and estimated $r_0$ using the Limber projection 
\citep[see][for details]{quadri07a}.  
The lower cutoff ensures that we are only measuring the large--scale 
correlation function, and are not unduly influenced by the clustering of 
galaxies within individual halos.  The upper cutoff was chosen to minimize 
the importance of the integral constraint correction \citep[see][]{quadri08}.  
The slope of the spatial correlation function is fixed to 1.6, which 
is consistent with each of the subsamples studied here, and the 
uncertainties are estimated using bootstrap resampling.  

\begin{figure}
\plotone{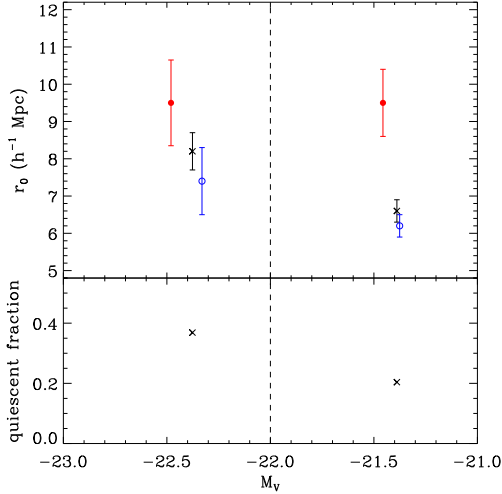}
\caption{Deprojected real--space clustering length in two absolute
magnitude bins ($M_V<-22$ and $M_V>-22$) for the $1.0<\zphot<2.0$ quiescent 
({\it red filled circles and errorbars}), star--forming 
({\it blue open circles}), and combined
({\it black crosses}) galaxy samples.  For each point, the median $M_V$
of the corresponding sample is plotted.  The lower panel shows the fraction
of quiescent galaxies in each bin.  Quiescent galaxies exhibit stronger
clustering, and thus occupy more massive dark matter halos, than star--forming
galaxies in both magnitude bins.
\label{fig_r0}}
\end{figure}

The resulting clustering lengths of the quiescent and
star--forming samples in both luminosity bins are shown in 
Figure~\ref{fig_r0}.  The clustering of bright
quiescent galaxies is stronger (at $1.4\sigma$ significance)
than that of the bright star--forming sample, with 
$r_0=9.5^{+1.1}_{-1.2}$\,h$^{-1}$\,Mpc and $7.4\pm 0.9$\,h$^{-1}$\,Mpc 
respectively, and the difference in clustering length becomes far more
pronounced in the fainter bin.\footnote{Since the objects which are ``missed''
in the faint luminosity bin are predominantly low--mass blue galaxies, the
true clustering of the star--forming sample in this bin is likely to be
even lower than what we find; i.e. the difference between $r_0$ for
the faint quiescent and star--forming galaxies would {\it increase} with
a deeper $K$ flux limit.}
Interestingly, the clustering length
of quiescent galaxies appears to be independent of luminosity; star--forming
galaxies, on the other hand, exhibit a marginal increase in clustering
strength with luminosity.   The stronger clustering of bright galaxies
in the {\it combined} (star--forming plus quiescent) sample thus appears to be
driven by both the larger quiescent fraction among bright galaxies as well
as the stronger clustering of star--forming galaxies.

The exact criteria used to divide the ``quiescent'' and ``star--forming''
samples (i.e. the lines shown in Figure~\ref{fig_evol}) may have an
effect on this result.  While we chose the diagonal dividing line to
lie centered between the star--forming track and quiescent ``clump,'' 
this is not necessarily the best criterion.  For example, moving the
line ``upward'' (toward redder $U-V$ and bluer $V-J$) would result in a
less complete sample of quiescent galaxies that is also less contaminated
by star--forming galaxies scattering across the line; similarly, moving
the line down toward the star--forming track increases completeness at
the expense of contamination.  To check whether this affects the clustering
measurement, we moved the diagonal divider
by $0.1$~magnitudes in both directions and re--calculated the $r_0$
values.  The perturbed $r_0$ values were all within $1\sigma$ of the
old values, with one exception: when the dividing line was moved upwards
(i.e. defining a less complete quiescent galaxy sample with less 
contamination),
the clustering strength of bright ($M_V<-22$) quiescent galaxies increased
by $\sim 1.5\sigma$.  Therefore, the result that quiescent galaxies
exhibit a larger clustering length appears robust to the color
criteria used to select the quiescent sample.

The clustering length we have found for quiescent galaxies
at $z=1-2$ is fully consistent with the value of $r_0=10.6\pm 1.6h^{-1}$\,Mpc 
for $\zphot=2-3$ DRGs in this catalog \citep{quadri08}.  This is perhaps
not surprising, as roughly half of the $\zphot>2$ DRGs in this sample
fall within the ``quiescent'' $UVJ$ selection region.

\section{Discussion} \label{sec_discussion}
The presence of a bimodal population of galaxies in the $UVJ$ plane 
to at least $z\sim 2$ is
consistent with previous spectroscopic studies, such as \citet{kriek08b},
who find a nascent red sequence at $z\sim 2.3$.
The purely photometric analysis presented herein allows
much larger samples of quiescent galaxies to be found, albeit with lower
confidence for any individual object.  At first glance, however, this
appears to be in conflict with the \citet{ciras07} claim that the
galaxy bimodality disappears at $z\sim 1.25-1.5$.  In fact, the 
apparent discrepancy
between these results is probably due to observational effects: \citet{ciras07}
consider only bimodality in the rest--frame color--magnitude ($U-B$ vs.~$M_B$) 
diagram, i.e.~the ``red sequence'' method.  Indeed, as the right panel
of Figure~\ref{fig_dahist} shows, such a plot does not show a bimodal
population of galaxies at $\zphot>1.5$, even though the bimodality is still
present in the $UVJ$ diagram.  Similarly, the absence of a bimodal distribution
in the $UVJ$ diagram at $z>2$ (Figure~\ref{fig_dahist}, left panel) does not
necessarily imply the disappearance of bimodality in the underlying galaxy
population, but rather reflects the larger photometric and $\zphot$
uncertainties at these redshifts (as can also be seen from the large $UVJ$
error bars at $2<\zphot<2.5$; Figure~\ref{fig_evol}).  

Using a single rest--frame color (determined via photometric redshifts)
to select quiescent and star--forming galaxies can
obscure an underlying bimodality for two reasons: (1) the population of
dusty, star--forming galaxies increases with redshift, and these have
significant overlap their optical colors with any red, quiescent galaxies that
may be present; and (2) the greater scatter in the rest--frame $U-V$ or $U-B$
at higher redshifts can effectively wash out the intrinsic red sequence.  This
increased scatter is due not only to the larger uncertainties on measured
photometry of fainter galaxies at large redshift, but also the
larger photometric redshift uncertainties \citep{kriek08a}.  
As previously mentioned, $U-B$ by itself is only effective at distinguishing 
quiescent from dusty galaxies when accurate spectroscopic redshifts are
available; with photometric redshifts the rest--frame color
uncertainty is too large to see the quiescent red sequence.  
The use of $V-J$ in addition to $U-V$ helps mitigate
both of these problems: $V-J$ separates red, quiescent galaxies from
red, dust--obscured starbursts (Figure~\ref{fig_rfcol}), and a 
two--dimensional color bimodality
is more resilient to increased scatter than a single color.  Thus, 
when only photometric redshifts are available, the two--color $UVJ$ technique 
employed herein appears to provide a ``cleaner''
selection of high--redshift, quiescent galaxies than standard red--sequence
methods.

The differences between high--redshift quiescent and star--forming
galaxy properties can provide clues to the physical mechanisms behind
star formation (or the lack thereof).  In the moderately high--redshift 
($z=1-2$) sample considered here, the most striking
contrasts between the two populations are in their clustering
strengths and luminosities.  Quiescent galaxies cluster more strongly than 
those undergoing active star formation (implying, from standard CDM theory,
that they reside preferentially in higher--mass halos) and typically
exhibit brighter absolute $V$ magnitudes, comprising about half of the bright
galaxy population at all redshifts.  However, the clustering strength
of quiescent galaxies does not appear to be strongly dependent on
luminosity: as shown in Figure~\ref{fig_r0}, the bright and faint 
quiescent samples exhibit roughly equal clustering lengths, which in turn
are consistently higher than the corresponding values for star--forming 
galaxies.  This may be indicative of a characteristic halo mass above which 
star formation is inhibited.

Our results may be taken to suggest that the star formation-density 
relation was already in place at $z\sim 1.5$, with quiescent galaxies
more strongly clustered and hence situated, on average, in denser regions.  
On first glance this would seem 
to contradict recent claims of a ``reversal'' of the 
star formation--density relation that occurs at $z\sim 1-1.5$ 
\citep[e.g.][]{elbaz07,cooper08}.  These authors present evidence 
that the average SFR per galaxy 
in dense regions is higher than in less dense regions, in contrast to the
well--known relationship in the local universe.  Whether this constitutes 
a ``reversal'' is in part a matter of terminology; it may be true that
the typical star formation rate increases with environmental density 
\citep[as found by, e.g.,][]{cooper08}, but this does not preclude the 
possibility that 
the relatively rare quiescent galaxies are preferentially found in the 
densest regions (as suggested by our results).  We also note that the 
star formation--density relation may be qualitatively different depending 
on whether one is considering absolute star formation rates or 
specific star formation rates \citep[see][]{cooper08}.

The observed segregation of quiescent and star--forming 
galaxies both in clustering strength and luminosity suggests
strong links between halo mass, stellar mass, and the cessation
of star formation.  Different theoretical bases for the quenching of star 
formation in galaxies hosted by massive halos have been proposed, 
some focusing on radio--mode AGN feedback processes which prevent
further gas cooling and star formation
\citep[e.g.][and references therein]{croton06,bower06}, while others
suggest that hot accretion and shock processes in high--mass halos
may be sufficient to produce quiescent galaxies without invoking AGN
\citep{keres05,birnboim07}.  Although these models
may not represent the definitive explanation for the observational results
presented here, the correspondence between theory and observation is
nonetheless intriguing.  Further observations of large quiescent galaxy
samples (e.g., comparing the radio emission of high--$z$ quiescent
and star--forming galaxies) are likely to shed further light on the underlying
quenching mechanism.

\section{Caveats} \label{sec_caveats}
This analysis employs photometric redshifts,
which of course are much less well--constrained than those determined
using spectroscopy.  Furthermore, the number of spectroscopic redshifts 
for comparison is quite small, with 119 available spec--$z$s in this field.
Essentially all of these known redshifts are from two samples, neither
of which can be considered unbiased: a set of
$z\sim 1$ old passively--evolving galaxies, and galaxies selected from
their radio emission (thus all likely exhibiting some degree of AGN
activity).  If anything, one might expect photometric redshift determinations
to be worse in this latter sample since the templates used to derive the
phot--$z$s do not include AGN activity; however, Figure~\ref{fig_zcompare} 
shows that the fits are actually quite good from about $z=0-1.5$
(and the worst outliers, including two--thirds those with signatures of AGN, 
are excluded through either a $\chi^2$ cut or the lack of a $\zphot$ 
solution).  The fit to the 
``old'' $z\sim 1$ galaxy sample was even better, with fractional redshift
errors on the order of 2\%.  At $z\ga 1.5$ it is likely that these fractional
errors are {\it at least} $\sim 0.07$, the scatter found in the EAZY--derived
photometric redshifts from the 
CDFS--GOODS catalog \citep[which included several more photometric bands
than the UDS$+$SXDS$+$SWIRE dataset;][]{brammer08}.

It is nonetheless possible that some additional systematic errors remain in the
photometric redshifts.  Such errors would manifest themselves as shifts
in the rest--frame colors, the magnitude of which primarily depend on the
galaxy redshift and template shape.  However, the results presented herein
are not dependent on the exact rest--frame colors themselves, but rather
on the relative separation between the quiescent and star--forming
populations.  Since the two populations are clearly visible from $z=0-2$
(Figure~\ref{fig_evol}), and the positions of galaxies in the $UVJ$ 
plane are strongly correlated with 
$24\mu$m flux (Figure~\ref{fig_div24um}), we conclude that
we can distinguish populations of galaxies based on their {\it relative}
star--formation rates.  It is nonetheless important to note that
the {\it absolute} values of the
interpolated rest--frame colors may still be subject to systematic,
possibly redshift--dependent offsets, and these offsets may be different
depending on the SED (e.g. for quiescent and star--forming
galaxies).  Further differences are likely 
to come about with differing photometric redshift codes and techniques
for determining rest--frame colors.  This underscores the importance
of defining such empirical color cuts based on the actual data and 
analysis techniques employed; relying on criteria defined on other datasets 
can lead to inaccurate results.

Another possible source of error is the ability of SExtractor to deblend
close pairs of objects.  Where two extended sources overlap, or
if the separation of two point sources is smaller than a few times
the $K$--band PSF FWHM ($\sim$0\farcs 7), oftentimes the two objects
will be counted as a single object.  In this case, the galaxy clustering
is probably underestimated at the smallest scales ($<1.5$\arcsec\ or so).  
Even if they are successfully identified
as separate objects, the fixed color apertures used to measure fluxes
are likely to be contaminated, and thus may cause errors in photometric
redshifts and rest--frame colors.  However, the fraction of objects in
such close pairs is negligible compared to the total catalog.  Furthermore,
these clustering results are based on larger--scale
correlations---i.e., we do not calculate the clustering below
$\theta\sim 2\farcs$.   Thus, this incompleteness at the smallest 
scales is unlikely to affect our results.

\section{Conclusions} \label{sec_conclusions}
From our matched multiband UDS$+$SXDS$+$SWIRE $K$--selected catalog we have 
derived reliable photometric redshifts and rest--frame colors.  For
a relatively bright ($K<22.4$) subsample of this catalog, we find that:
\begin{enumerate}
\item{Galaxies show strong bimodal behavior in rest--frame $U-V$ vs.
$V-J$ color--color space, with one population of ``dead'' (non--star
forming) galaxies, and a sequence of dusty, actively star--forming
galaxies.  This behavior can be seen out to $z\sim 2$, so the observed 
present--day galaxy bimodality was present at least to this redshift;
however, at $z\ga 1.5$ the bimodality is not seen in a single--color--magnitude
diagram.}
\item{From $z=0$ to $2.5$, the rest--frame $V-J$ color (at red $U-V$ colors) 
shows a strong correlation with our estimates of specific star--formation
rate, indicating that
the $V-J$ color is a reliable tracer of star formation activity even at 
$z>2$ where the rest--frame color uncertainties are too large for the
bimodality to be seen.}
\item{Quiescent and star--forming galaxies at $z>1$ contribute roughly equally
to the overall galaxy population at the brightest $V$ luminosities;
the less luminous population is dominated by actively star--forming objects.}
\item{The clustering strength of quiescent galaxies from $z=1-2$ 
appears to be independent of galaxy luminosity, and is consistently
stronger than the clustering of the actively star--forming galaxies;
thus, ``dead'' galaxies appear
to occupy more massive halos than those which are in the process of 
forming stars.  This suggests a link
between halo mass and the early cessation of star formation activity.}
\end{enumerate}

Forthcoming data releases of the UKIDSS UDS, including deep 
$H$--band data
and ultimately reaching a depth of $K=24.9$ (AB), will allow these results
to be tested with even fainter, higher--redshift galaxies.  More importantly,
the 290--hour ultra--deep {\it Spitzer} legacy survey of the UDS field 
({\it SpUDS}; PI: J.~Dunlop) currently being undertaken will provide
at least an order--of--magnitude increase in exposure time compared
to the SWIRE data used in this 
paper, giving much--improved constraints on photometric redshifts,
AGN activity, and rest--frame colors.  

\acknowledgments
We are grateful to the UKIDSS, SXDS, and Spitzer/SWIRE teams for making
their reduced data available, and Chris Simpson for providing us with
his spectroscopic redshifts in this field.  We also thank Gabe Brammer and Ned
Taylor for their extensive help with computing photometric redshifts
and rest--frame colors respectively, Stijn Wuyts for providing the 
specific star--formation rates from {\it FIREWORKS}, and Mariska Kriek 
and the anonymous referee for many constructive comments and suggestions.
R.J.W. acknowledges the support of the Netherlands Organization for Scientific
Research (NWO) and the Leids Kerkhoven--Bosscha Fonds.  R.Q. is 
supported by a NOVA postdoctoral fellowship.
This research has made use of the NASA/IPAC Extragalactic Database (NED) 
which is operated by the Jet Propulsion Laboratory, California Institute of 
Technology, under contract with the National Aeronautics and Space 
Administration.

\appendix
\section{Estimating specific star--formation rates} \label{sec_appendix}
Accurately determining specific star formation rates (SSFR; i.e. SFR as a 
fraction of galaxy mass) typically requires deep UV and $24\mu$m
data in order to measure unobscured and obscured star formation, respectively.
Unfortunately, the UDS data presented here currently lack several
key ingredients; i.e. there are no $U$-band data, the available
{\it SWIRE} $24\mu$m data are shallow, and the necessary detailed
models for determining galaxy masses and the $24\mu$m--$L_{IR}$ conversions
are beyond the scope of this work.  However, to test the validity
of the $UVJ$ colors as a tracer of SSFR, such precision is not necessary:
simple order--of--magnitude estimates will suffice.

Over the redshift range $0.5<z<2.5$, we therefore resort
to a proxy based entirely on observed photometry and calibrated with
the Chandra Deep Field--South {\it FIREWORKS} catalog of \citet{wuyts08}.  
Reliable SSFRs were
derived from {\it FIREWORKS} based on the rest--frame 2800\AA\ 
luminosity and observed $24\mu$m fluxes (Wuyts et al., in preparation).
The sum of these two contributions should then
approximate the total amount of star formation in ``typical'' galaxies
(i.e. those without extreme obscuration, like SMGs).  Since $K$ flux
is itself a tracer of stellar mass, it follows that the
$f_{\rm UV}/f_K$ flux ratio (for some observed band that falls in or
near the rest--frame ultraviolet) can provide a rough
estimate of the unobscured SSFR, and $f_{24\mu m}/f_K$ should do the
same for the ``dusty'' SSFR.

\begin{figure}
\epsscale{0.6}
\plotone{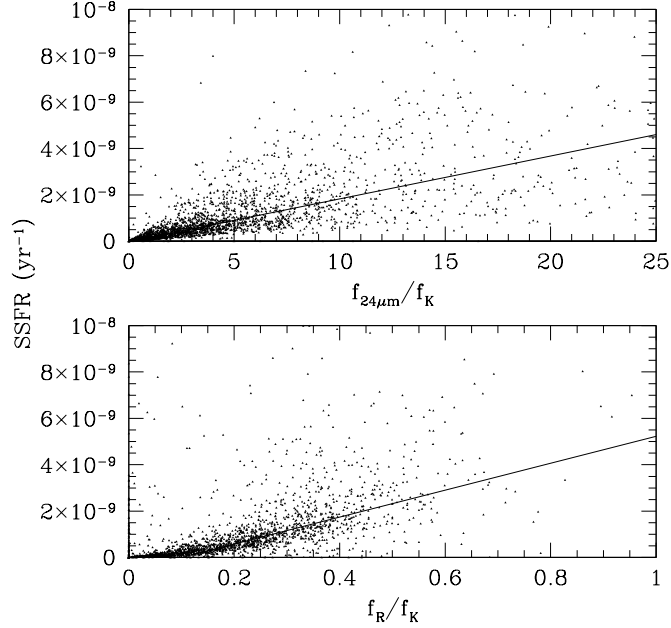}
\caption{Relations used to derive specific star--formation rates
from the observed IR ($f_{\rm 24\mu m}/f_K$; {\emph top panel}) and UV
($f_R/f_K$; {\emph bottom panel}) flux ratios.  Data in this figure
are from the FIREWORKS catalog of \citet{wuyts08}; the IR relation shown is fit to 
galaxies between $0.5<z<2.5$ and the UV at $1.0<z<2.5$ (a similar UV
relation was derived for the $z=0.5-1$ galaxies, not shown here). 
The IR SSFR shows a good correlation with $f_{24\mu m}/f_K$ over the full 
range, but the UV relation tends to underestimate large SSFRs.
\label{fig_ssfrrel} }
\end{figure}

From the {\it FIREWORKS} data we find that a single linear relation sufficiently
describes the correlation between the IR SSFR and $f_{24\mu m}/f_K$ 
over $0.5<z<2.5$.  There is also a strong correlation between
UV SSFR and $f_R/f_K$; however, the exact slope and normalization of the
relation appear to change abruptly at $z\sim 1$; for the UV relation
we thus perform separate linear fits at $0.5<z<1$ and $1<z<2.5$.
This UV proxy is somewhat flatter (i.e. insensitive to $f_R/f_K$) than
the $24\mu$m--SFR relation, and substantially underestimates the
largest UV SSFRs by a factor of $\sim 2$, but it appears to accurately
reproduce relatively low SSFRs ($la 2\times 10^9$\,yr$^{-1}$); thus, this UV 
proxy is effectively used as an additive term to correct the fluxes of galaxies
which are faint in $24\mu$m.  The best--fit relations, shown in
Figure~\ref{fig_ssfrrel}, are:
\begin{equation}
\begin{array}{l}
{\rm SSFR}_{\rm IR} = -2.2\times 10^{-11} + 1.85\times 10^{-10} (f_{24}/f_K)\\
{\rm SSFR}_{\rm UV} (z<1) = -7.4\times 10^{-10} + 3.8\times 10^{-9} (f_R/f_K)\\
{\rm SSFR}_{\rm UV} (z>1) = -5.8\times 10^{-10} + 5.8\times 10^{-9} (f_R/f_K)
\end{array}
\end{equation}
Note that the UV relations formally yield substantially negative SSFR values
even when $f_R/f_K>0$, since SSFR$_{\rm UV}$ is not quite a linear function
of $f_R/f_K$; thus, we require that
the UV contribution to the total SSFR be positive (i.e. for any object
where SSFR$_{\rm UV}<0$, we set SSFR$_{\rm UV}=0$).  Since a large number
of galaxies have $f_{24\mu m}<0$ due to background noise, and the 
SSFR$_{\rm IR}$ relation appears very close to linear,
we do allow individual objects to have SSFR$_{\rm IR}<0$ in the stacking 
analysis.

\begin{thebibliography}{}
\bibitem[Baldry et al.(2004)]{baldry04} Baldry, I.~K., Glazebrook, K.,
         Brinkmann, J., Ivezi\'c, \v{Z}., Lupton, R~.H., Nichol, R.~C.,
         \& Szalay, A.~S.~2004, \apj, 600, 681
\bibitem[Bertin \& Arnouts(1996)]{bertin96} Bertin, E., \& Arnouts, S.~1996,
         A\&AS, 117, 393
\bibitem[Bertin et al.(2002)]{bertin02} Bertin, E., Mellier, Y., Radovich, M.,
         Missonnier, G., Didelon, P., \& Morin, B.~2002, in ASP Conf. Ser.
	 281, Astronomical Data Analysis Software and Systems XI, ed. D.~A.
	 Bohlender, D.~Durand, \& T.~H.~Handley (San Francisco: ASP), 228
\bibitem[Bessell(1990)]{bessell90} Bessell, M.~S.~1990, \pasp, 102, 1181
\bibitem[Birnboim, Dekel, \& Neistein(2007)]{birnboim07} Birnboim, Y., Dekel, 
         A., \& Neistein, E.~2007, \mnras, 380, 339
\bibitem[Blanton et al.(2003)]{blanton03} Blanton, M.~R., et al.~2003,
         \apj, 592, 819
\bibitem[Bolzonella, Miralles, \& Pell\'o(2000)]{hyperz} Bolzonella,
         M., Miralles, J.-M., \& Pell\'o, R.~2000, \aap, 363, 476
\bibitem[Bower et al.(2006)]{bower06} Bower, R.~G., et al.~2006,
         \mnras, 370, 645
\bibitem[Brammer et al.(2008)]{brammer08} Brammer, G.~B., van Dokkum,
         P.~G., \& Coppi, P.~2008, \apj, 686, 1503
\bibitem[Bruzual \& Charlot(2003)]{bc03} Bruzual, G. \& Charlot, S.~2003,
         \mnras, 344, 1000
\bibitem[Budavari et al.(2003)]{budavari03} Budav\'ari, T., et al.~2003,
         \apj, 595, 59
\bibitem[Casali et al.(2007)]{casali07} Casali, M., et al.~2007, \aap,
         467, 777
\bibitem[Cirasuolo et al.(2007)]{ciras07} Cirasuolo, M., et al.~2007,
         \mnras, 380, 585
\bibitem[Cooper et al.(2008)]{cooper08} Cooper, M.~C., et al.~2008, 
         \mnras, 383, 1058
\bibitem[Croton et al.(2006)]{croton06} Croton, D., et al.~2006, 
         \mnras, 365, 11
\bibitem[Daddi et al.(2004)]{daddi04} Daddi, E., et al.~2004, \apj, 617, 746
\bibitem[Daddi et al.(2005)]{daddi05} Daddi, E., et al.~2005, \apj, 626, 680
\bibitem[Elbaz et al.(2007)]{elbaz07} Elbaz, D., et al.~2007, \aap, 468, 33
\bibitem[F\"orster Schreiber et al.(2006)]{forster06} F\"orster Schreiber,
         N., et al.~2006, \aj, 131, 1891
\bibitem[Franx et al.(2003)]{franx03} Franx, M., et al.~2003, \apj, 587, L79
\bibitem[Gawiser et al.(2006)]{gawiser06} Gawiser, E., et al.~2006, \apjs,
         162, 1
\bibitem[Giavalisco et al.(2004)]{giavalisco04} Giavalisco, M., et al.~2004,
         \apj, 600, L93
\bibitem[Grazian et al.(2007)]{grazian07} Grazian, A., et al.~2007, \aap,
         465, 393
\bibitem[Hambly et al.(2008)]{hambly08} Hambly, N.~C., et al.~2008,
         \mnras, 384, 637
\bibitem[Hewett et al.(2006)]{hewett06} Hewett, P.~C., Warren, S.~J., 
         Leggett, S.~K., \& Hodgkin, S.~T.~2006, \mnras, 367, 454
\bibitem[Kauffmann et al.(2003)]{kauffmann03} Kauffmann, G., et al.~2003,
         \mnras, 341, 33
\bibitem[Keres et al.(2005)]{keres05} Keres, D., Katz, N., Weinberg,
         D.~H., \& Dav\'e, R.~2005, \mnras, 363, 2
\bibitem[Kriek et al.(2006)]{kriek06} Kriek, M., et al.~2006, \apj, 649, L71
\bibitem[Kriek et al.(2008a)]{kriek08a} Kriek, M., et al.~2008, \apj, 677, 219
\bibitem[Kriek et al.(2008b)]{kriek08b} Kriek, M., et al.~2008, \apj, 682, 896
         (arXiv:0804.4175)
\bibitem[Kron(1980)]{kron80} Kron, R.~1980, \apj, 43, 305
\bibitem[Labb\'e et al.(2003)]{labbe03} Labb\'e, I., et al.~2003, \aj, 125, 1107
\bibitem[Labb\'e et al.(2005)]{labbe05} Labb\'e, I., et al.~2005, \apj, 
         624, L81
\bibitem[Labb\'e et al.(2006)]{labbe06} Labb\'e, I., Bouwens, R., 
         Illingworth, G.~D., \& Franx, M.~2006, \apj, 649, L67
\bibitem[Landy \& Szalay(1993)]{landy93} Landy, S.~D., \& Szalay, A.~S.~1993,
         \apj, 412, 64
\bibitem[Lawrence et al.(2007)]{lawrence07} Lawrence, A., et al.~2007,
         \mnras, 379, 1599
\bibitem[Lonsdale et al.(2003)]{lonsdale03} Lonsdale, C.~J., et al.~2003, 
         \pasp, 115, 897
\bibitem[Miyazaki et al.(2002)]{miyazaki02} Miyazaki, S., et al.~2002, 
         \pasj, 54, 833
\bibitem[Papovich et al.(2006)]{papovich06} Papovich, C., et al.~2006,
         \apj, 640, 92
\bibitem[Quadri et al.(2007a)]{quadri07a} Quadri, R., et al.~2007a, \apj,
         654, 138
\bibitem[Quadri et al.(2007b)]{quadri07b} Quadri, R., et al.~2007b, \aj, 
         134, 1103
\bibitem[Quadri et al.(2008)]{quadri08} Quadri, R., et al.~2008, \apjl, 
         685, 1
\bibitem[Rudnick et al.(2003)]{rudnick03} Rudnick, G., et al.~2003,
         \apj, 599, 847
\bibitem[Rudnick et al.(2006)]{rudnick06} Rudnick, G., et al.~2006,
         \apj, 650, 624
\bibitem[Sekiguchi et al.(2004)]{sekiguchi04} Sekiguchi, K., et al.~2004,
         BAAS, 36, 1478
\bibitem[Simpson et al.(2006)]{simpson06} Simpson, C., et al.~2006, 
         \mnras, 372, 741
\bibitem[Steidel et al.(1996)]{steidel96} Steidel, C.~C., Giavalisco, M., 
         Pettini, M., Dickinson, M., Adelberger, K.~L.~1996, \apj, 462, L17
\bibitem[Stern et al.(2005)]{stern05} Stern, D., et al.~2005, \apj, 631, 163
\bibitem[Tokunaga et al.(2002)]{tokunaga02} Tokunaga, A.~T., Simons, D.~A.,
         \& Vacca, W.~D.~2002, \pasp, 114, 180
\bibitem[van Dokkum et al.(2006)]{vdokkum06} van Dokkum, P.~G., et al.~2006,
         \apj, 638, L59
\bibitem[Warren et al.(2007a)]{warren07a} Warren, S.~J., et al.~2007,
         \mnras, 375, 213
\bibitem[Warren et al.(2007b)]{warren07b} Warren, S.~J., et al.~2007,
         preprint, arXiv:astro-ph/0703037
\bibitem[Wuyts et al.(2007)]{wuyts07} Wuyts, S., et al.~2007, \apj, 655, 51
\bibitem[Wuyts et al.(2008)]{wuyts08} Wuyts, S., Labb\'e, I., F\"orster
         Schreiber, N., Franx, M., Rudnick, G., Brammer, G., \& van
         Dokkum, P.~2008, \apj, 682, 985
\bibitem[Yamada et al.(2005)]{yamada05} Yamada, T., et al.~2005, \apj, 634, 861
\bibitem[Yan et al.(2004)]{yan04} Yan, L., et al.~2004, \apjs, 154, 75
\end{thebibliography}
\end{document}